\documentclass[%
 reprint,
superscriptaddress,
 amsmath,amssymb,
 aps,
]{revtex4-2}

\usepackage{graphicx}
\usepackage{multirow}
\usepackage{physics}
\usepackage{tcolorbox}
\usepackage{booktabs}
\usepackage{quantikz}
\usepackage{hyperref}
\usepackage{color}
\usepackage{array}
\usepackage[normalem]{ulem}

\begin{document}

\title{Unfolded distillation: very low-cost magic state preparation for biased-noise qubits}

\author{Diego Ruiz}
\email{diego.ruiz@alice-bob.com}%
\affiliation{Alice \& Bob, 49 Bd du Général Martial Valin, 75015 Paris, France}
\affiliation{Laboratoire de Physique de l'École Normale Supérieure, École Normale Supérieure, Centre Automatique et Systèmes, Mines Paris, Université PSL, CNRS, Inria, Paris, France}
\author{J\'erémie Guillaud}%
\affiliation{Alice \& Bob, 49 Bd du Général Martial Valin, 75015 Paris, France}
\author{Christophe Vuillot}%
\affiliation{Alice \& Bob, 49 Bd du Général Martial Valin, 75015 Paris, France}
\author{Mazyar Mirrahimi}%
\affiliation{Laboratoire de Physique de l'École Normale Supérieure, École Normale Supérieure, Centre Automatique et Systèmes, Mines Paris, Université PSL, CNRS, Inria, Paris, France}

\date{\today}

\begin{abstract}
Magic state distillation enables universal fault-tolerant quantum computation by implementing non-Clifford gates via the preparation of high-fidelity magic states. However, it comes at the cost of substantial logical-level overhead in both space and time. In this work, we propose a very low-cost magic state distillation scheme for biased-noise qubits. By leveraging the noise bias, our scheme enables the preparation of a magic state with a logical error rate of $3 \times 10^{-7}$, using only 53 qubits and 5.5 error correction rounds, under a noise bias of $\eta \gtrsim 5 \times 10^6$ and a phase-flip noise rate of 0.1\%. This reduces the circuit volume by more than one order of magnitude relative to magic state cultivation for unbiased-noise qubits and by more than two orders of magnitude relative to standard magic state distillation. Moreover, our scheme provides three key advantages over previous proposals for biased-noise qubits. First, it only requires nearest-neighbor two-qubit gates on a 2D lattice. Second, the logical fidelity remains nearly identical even at a more modest noise bias of $\eta \gtrsim 80$, at the cost of a slightly increased circuit volume. Third, the scheme remains effective even at high physical phase-flip rates, in contrast to previously proposed approaches whose circuit volume grows exponentially with the error rate. Our construction is based on unfolding the $X$ stabilizer group of the Hadamard 3D quantum Reed-Muller code in 2D, enabling distillation at the physical level rather than the logical level, and is therefore referred to as \textit{unfolded} distillation.
\end{abstract}

\maketitle

Quantum computing promises to solve certain computational problems intractable for classical computers~\cite{shor1994algorithms,dalzell2023quantum}, but is hindered by errors which prevent performing long computations. Although the first useful applications would typically require an error rate below $10^{-6}$~\cite{beverland2022assessing}, the best quantum platforms have an error rate of the order of $10^{-2}-10^{-3}$~\cite{acharya2024quantum,moses2023race,bluvstein2024logical}.

Fault-tolerant quantum computation offers a solution to this problem~\cite{shor1996fault}. First, the logical qubits processed during the algorithm are encoded in a quantum error correcting code across a collection of physical qubits, ensuring robustness against errors through redundancy~\cite{shor1995scheme,steane1996multiple}. If the physical error rate is below a certain threshold, the logical error rate can be exponentially reduced by increasing the distance $d$ of the code~\cite{aharonov1997fault,kitaev2003fault,knill1998resilient}. Then, logical gates, used to perform the quantum algorithm, must be implemented in a fault-tolerant manner to avoid compromising the protection brought by the error correcting code. For instance, transversal gates, which consist of applying physical gates independently to each physical qubit, are naturally fault-tolerant because physical qubits of the same logical codeblock do not interact.

\begin{figure}[ht]
    \centering
    \includegraphics[width=0.4\textwidth]{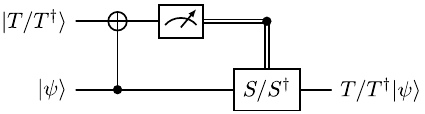}
    \caption{Quantum gate teleportation using a magic state~\cite{zhou2000methodology}. A $\ket{T/T^\dagger} = \tfrac{1}{\sqrt{2}}(\ket{0} + e^{\pm i \pi/4} \ket{1})$ magic state is consumed via a quantum teleportation circuit to implement a $T/T^\dagger$ gate on an arbitrary input state $\ket{\psi}$. This can be used either to implement a $T/T^\dagger$ gate on a logical qubit within an algorithm, provided a high-fidelity $\ket{T/T^\dagger}$ state, or to apply a noisy $T/T^\dagger$ gate using a noisy $\ket{T/T^\dagger}$ state as part of the magic state distillation process~\cite{bravyi2005universal}.}
    \label{fig:T_teleport}
\end{figure}

Nevertheless, the Eastin-Knill theorem states that, for a given quantum error correcting code, transversal gates alone cannot form a universal set of gates~\cite{eastin2009restrictions}. Moreover, it has been proven that, for topological codes in 2D, which are relevant for most platforms due to connectivity constraints, only the set of Clifford gates can be implemented transversally~\cite{bravyi2013classification}. Hence, at least one non-Clifford gate must be implemented by a different means. One of the most studied schemes involves preparing a high-fidelity logical $\ket{T} = \tfrac{1}{\sqrt{2}}(\ket{0} + e^{i \pi/4} \ket{1})$ magic state~\cite{bravyi2005universal}. The magic state is then consumed in a quantum teleportation circuit to implement the $T = R_Z(\pi/4)$ gate on any logical qubit, as shown in Figure~\ref{fig:T_teleport}~\cite{zhou2000methodology}.

\begin{table*}
\centering
\renewcommand{\arraystretch}{1.2}
\begin{tabular}{c!{\vrule width 1pt}c|c|c!{\vrule width 1pt}c|c|c|c}
 & Physical gate & \multirow{2}{*}{Physical error} & \multirow{2}{*}{Noise bias} & \multirow{2}{*}{Logical error} & Number & Number & \multirow{2}{*}{Qubitcycles}   \\ 
 & required & & & & of qubits &  of cycles & \\
\hline
\multicolumn{8}{c}{\color{white}........................................... \color{black}Biased-noise Toffoli schemes} \\
\hline
\hline
\multirow{2}{*}{Bottom-up~\cite{chamberland2022building,gouzien2023performance}} & \multirow{2}{*}{CC$X$} & $p_Z = 10^{-3}$ & $5 \times 10^7$  & $5\times 10^{-8}$ & 46
 & 12.4 & 570  \\ 
 &  & $p_Z = 5 \times 10^{-3}$ & $5 \times 10^5$  & $3 \times 10^{-5}$ & 46 & 1380 & 63\ \!500  \\ 
\hline
\multicolumn{8}{c}{\color{white}........................................... \color{black}Unbiased-noise schemes} \\
\hline
\hline
Magic state distillation~\cite{litinski2019magic}& $T$ & $p=10^{-3}$ & 1& $4.5 \times 10^{-8}$ & 4\ \!620 & 42.6 & 197\ \!000  \\ 
\hline
\multirow{2}{*}{Magic state cultivation~\cite{gidney2024magic}}& \multirow{2}{*}{$T$} & \multirow{2}{*}{$p=10^{-3}$} & \multirow{2}{*}{1}& $9 \times 10^{-10}$ & $463$ & $216$ & 100\ \!000 \\
 &  &  & & $7 \times 10^{-7}$ & $463$ &28 & 13\ \!000   \\  
 \hline
\multicolumn{8}{c}{\color{white}............................................ \color{black}\textbf{Unfolded distillation (this work)}} \\
\hline
\hline
\multirow{4}{*}{\shortstack{High noise bias with\\ Repetition code}} & \multirow{4}{*}{$X^{1/4}$} & \multirow{2}{*}{$p_Z=10^{-3}$} & $5 \times 10^6$& $3 \times 10^{-7}$ & 53 & 5.5 & 292   \\
 & & & $5 \times 10^7$ & $6 \times 10^{-8}$ & 53 & 12 & 636   \\
 \cline{3-8}
 & & \multirow{2}{*}{$p_Z=5 \times10^{-3}$} & $10^6$ & $4 \times 10^{-5}$ & 57 & 17.7 & 1\ \!010   \\
 & & & $3 \times 10^6$ & $2 \times 10^{-5}$ & 57 & 15 & 855   \\
\hline
Modest noise bias with & \multirow{2}{*}{$X^{1/4}$} & \multirow{2}{*}{$p_Z=10^{-3}$} & \multirow{2}{*}{80} & \multirow{2}{*}{$7 \times 10^{-7}$} & \multirow{2}{*}{175} & \multirow{2}{*}{9.6} & \multirow{2}{*}{1\ \!680}  \\
Surface code & & & & & & &  \\
 \hline
\end{tabular}
\caption{Magic state preparation overhead comparison. The table highlights both the hardware requirements and the performance of various schemes. The biased-noise Toffoli schemes~\cite{guillaud2021error, chamberland2022building, gouzien2023performance}, which prepare a Toffoli magic state (see Section~\ref{sec:state of the art}B), while being efficient at a low physical error rate, require a significant noise bias and perform poorly at high physical error rates. For unbiased-noise qubits, the magic state distillation protocol~\cite{litinski2019magic}, while robust against high physical error rates, entails substantial overhead (see Section~\ref{sec:state of the art}A). Conversely, the magic state cultivation protocol~\cite{gidney2024magic} is less resource-intensive but highly sensitive to the physical error rate (see Section~\ref{sec:state of the art}B). \textit{Unfolded} distillation, the protocol proposed in this work, achieves very low overhead while retaining the main benefits of conventional magic state distillation. It remains effective under high physical error rates and requires only two-qubit gates, unlike biased-noise Toffoli schemes. Moreover, the scheme can be adapted to a moderate noise bias of $\eta \approx 80$, at the expense of a slight increase in the number of qubits. The first and third entries of the unfolded distillation using repetition codes correspond to the scheme presented in Section~\ref{sec:repetition code}, while the second and fourth entries correspond to the scheme in Appendix~\ref{app:Clifford ladder}. The surface code version of the scheme is presented in Section~\ref{sec:surface_code}. In both the biased-noise Toffoli schemes and unfolded distillation with repetition codes, the noise bias is chosen such that the logical bit-flip error rate is approximately an order of magnitude lower than the logical phase-flip error rate. Appendix~\ref{app:architecture comparison} details how all values in this table were obtained.}
\label{tab:overhead}
\end{table*}

However, preparing a magic state with a high fidelity is costly. The most studied approach, known as magic state distillation, takes multiple noisy states as input and produces fewer but higher-fidelity states~\cite{bravyi2005universal}. Its high cost arises from the need to perform the distillation protocol at the logical level, introducing significant space and time overhead, even though remarkable progress has been made in reducing its cost since the initial proposal~\cite{litinski2019magic,gidney2019efficient}. In contrast to distillation-based methods, schemes relying on the measurement of a Clifford operator~\cite{shor1996fault,gottesman1999demonstrating,zhou2000methodology} have recently been shown to outperform distillation for a physical error rate of $p=10^{-3}$, up to a target logical error of $10^{-9}$~\cite{gidney2024magic} (see Table~\ref{tab:overhead}, entry \textit{'Unbiased-noise schemes'}). While the space overhead of Clifford measurement schemes is minimal, they still require a large time overhead, which increases exponentially with the physical error rate due to postselection.

Despite these substantial advances in magic state preparation, the overhead of fault-tolerance remains a significant challenge. An avenue that has been explored to reduce this cost is the use of biased-noise qubits, where errors with a bit-flip component $X$ and $Y$ are suppressed compared to the phase-flip $Z$ errors ($p_X,p_Y \ll p_Z$). Such a noise structure can be observed in spin qubits that naturally exhibit a noise bias~\cite{steinacker2024300}. It has been shown that this property can be exploited to increase the error-correction threshold by a factor of 2 to 4~\cite{aliferis2008fault}, by additionally leveraging the fact that the two-qubit $ZZ$ interaction is bias-preserving, \textit{i.e.} does not induce bit-flip errors, in spin and superconducting qubits~\cite{taylor2005fault,brito2008efficient}. Furthermore, it was found that when the correlated phase-flip error associated with the $ZZ$ interaction occurs at a lower rate than single-qubit $Z$ errors, this asymmetry can be exploited to facilitate the preparation of a $\ket{T}$ magic state~\cite{webster2015reducing}. A key development, however, was the proposal to encode qubits in harmonic oscillators in a way that yields an extremely large noise bias while also supporting bias-preserving CNOT and Toffoli gates~\cite{mirrahimi2014dynamically,guillaud2019repetition,puri2020bias}. Cat qubits implemented in superconducting circuits, which encode quantum information in coherent states of opposite phases within a harmonic oscillator, can exhibit a significant noise bias of $\eta = p_Z / (p_X + p_Y) \gtrsim 10^7$~\cite{reglade2024quantum}. Recent experiments have reported a macroscopic bit-flip lifetime of 22 seconds and a phase-flip time of $1.3 \times 10^{-6}$ seconds~\cite{rousseau2025enhancing}. Alternatively, theoretical proposals have been made to implement cat qubits in ion traps~\cite{rojkov2024stabilization} and neutral atoms~\cite{omanakuttan2024fault}, along with some preliminary experimental results~\cite{debry2025error}. If the noise bias is sufficiently large, and the physical CNOT operations (required for error syndrome measurements) can be performed while preserving this bias, the quantum information can be encoded in a phase-flip repetition or LDPC code~\cite{ruiz2025ldpc}, substantially reducing the physical qubit overhead required for quantum error correction~\cite{chamberland2022building,gouzien2023performance}. This was demonstrated in a recent experiment, where cat qubits concatenated into a phase-flip repetition code were operated below the phase-flip threshold~\cite{putterman2025hardware}. Alternatively, if bit-flip errors are not sufficiently suppressed relative to the quantum volume of the algorithm, a thin surface code~\cite{hann2024hybrid} or an XZZX code~\cite{bonilla2021xzzx} with $d_X \le d_Z$ could be used compared to a square patch of surface code.

But if biased-noise qubits can strongly reduce the cost of a quantum memory, it is however less clear how much additional benefit could optimally be expected for magic state preparation. Magic state distillation could be straightforwardly implemented by substituting the surface code logical qubits with codes tailored for the noise bias. While it has been shown that the raw input magic states could be prepared with a better fidelity~\cite{singh2022high}, it would still incur significant overhead in both time and space (see Table IV in Ref.~\cite{chamberland2022building}). Alternatively, several works have focused on Clifford measurement schemes~\cite{chamberland2022building,gouzien2023performance}, yet this method presents several drawbacks. First, they are based on physical Toffoli gates, requiring a challenging coupling between three physical qubits. Second, while these schemes have a very small physical qubit overhead, they require very low physical error rates for the time overhead to remain manageable, as is also the case with magic state cultivation~\cite{gidney2024magic}, due to the use of postselection. Third, a significant noise bias must be preserved during the execution of the Toffoli gate, a requirement that has not yet been demonstrated. The performances of these schemes are summarized in the Section \textit{'Biased-noise Toffoli schemes'} of Table~\ref{tab:overhead}.

In this work, we propose a very low-cost magic state preparation scheme for biased-noise qubits. At a physical error rate of $p_Z = 10^{-3}$ and a noise bias of $\eta \gtrsim 5 \times 10^6$, the scheme outputs a magic state with a logical error rate of $3 \times 10^{-7}$ with only 53 physical qubits and less than $6$ cycles of error correction. Moreover, our construction offers three key advantages over previously proposed schemes. First, unlike schemes that rely on the Toffoli gate, our scheme only requires 2-qubit bias-preserving CNOT gates and nearest-neighbor connectivity in a 2D planar architecture. Second, it remains effective under a more modest noise bias while requiring only a slightly larger number of qubits. Under a noise bias of $\eta = 80$ and with a physical error rate of $p_Z = 10^{-3}$, the scheme outputs a magic state with a logical error rate of $7 \times 10^{-7}$ with 175 physical qubits and less than $10$ cycles of error correction. Third, it remains effective even at higher physical error rates, a regime in which Toffoli gate based schemes perform poorly. At a physical error rate of $p_Z = 5 \times 10^{-3}$, and for the same logical error rate of $2\times 10^{-5} - 3\times 10^{-5}$, the \textit{Bottom-up} Toffoli protocol~\cite{chamberland2022building,gouzien2023performance} requires approximately 75 times more qubitcycles. Table~\ref{tab:overhead} compares our construction with schemes proposed for magic state preparation for both biased and standard (unbiased-noise) qubits.

The key insight of our construction is that the noise bias structure enables magic state distillation to be performed directly at the physical level rather than the logical level. Specifically, our construction relies on the unfolding of the X-type stabilizer group of the Hadamard 3D quantum Reed-Muller code~\cite{knill1998resilient, bravyi2005universal} into a 2D nearest-neighbor layout. This technique, which we refer to as \textit{unfolded} distillation, as it operates at the physical level, achieves extremely low overhead.

The manuscript is organized as follows. In Section~\ref{sec:state of the art}, we review the state of the art of magic state preparation, focusing on the two leading methods: magic state distillation and Clifford measurement. Section~\ref{sec:main_ideas} introduces the main ideas behind our novel construction. In Section~\ref{sec:repetition code}, we provide more details about the protocol and detail the code construction. Section~\ref{sec:surface_code} discusses how our scheme can be adapted to a more modest noise bias. Section~\ref{sec:universal} presents the construction of a universal fault-tolerant gate set based on unfolded distillation, which also enables efficient preparation of $\ket{S} = \ket{+i}$, $\ket{CZ} = \ket{00} + \ket{01} + \ket{10} - \ket{11}$ states and $\ket{C_XC_XX} = C_XC_XX\ket{000}$ magic states, where $C_XC_X X = (H \otimes H \otimes I) CCX (H \otimes H \otimes I)$. Finally, Section~\ref{sec:hybrid} discusses an implementation of unfolded distillation on superconducting cat qubits. 

\section{State of the art}
\label{sec:state of the art}

\subsection{Distillation}

\begin{figure}[t]
    \centering
    \includegraphics[width=0.5\textwidth]{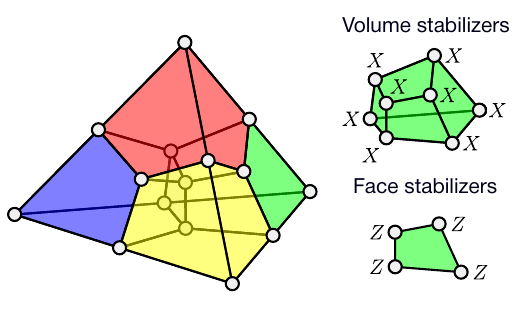}
    \caption{Quantum Reed-Muller code~\cite{steane2002quantum}. The $[[15,1,3]]$ quantum Reed-Muller code encodes one logical qubit into 15 physical qubits with distance $d = 3$. The tetrahedron is divided into four hexahedrons. The volume of each hexahedron represents an $X$-type stabilizer of weight 8, while the faces correspond to $Z$-type stabilizers of weight 4. This is the smallest distance-3 code that admits a transversal $T$ gate~\cite{koutsioumpas2022smallest} and is commonly used in magic state distillation schemes~\cite{bravyi2005universal}.}
    \label{fig:Reed-Muller}
\end{figure}

Magic state distillation is one of the leading approaches for preparing high-fidelity magic states~\cite{bravyi2005universal}. At a high level, the protocol takes low-fidelity magic states as input and produces fewer magic states as output, but with higher fidelities. Magic state distillation operates at the logical level, and thus requires input states to be encoded in a quantum error correcting code. This can be done using state injection, which typically results in a logical magic state whose fidelity is comparable to the physical error rate~\cite{li2015magic,gidney2023cleaner}. Importantly, with the exception of the input magic states, all logical gates in the magic state distillation circuit are Clifford gates, which can be executed with high fidelity at a lower cost than non-Clifford gates.

\begin{figure}[h]
    \centering
    \includegraphics[width=0.3\textwidth]{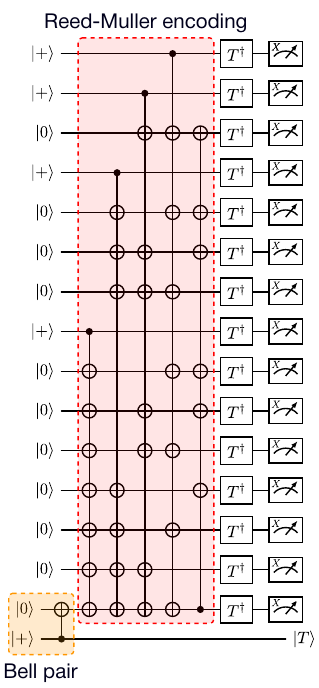}
    \caption{Magic state distillation protocol~\cite{fowler2012surface}. The protocol uses 16 logical qubits, typically implemented as surface code patches. It begins by creating a logical Bell pair between two logical qubits, one of which is then encoded with 14 additional logical qubits using the Reed-Muller code.  A transversal $T$ gate is applied to all qubits in the encoded block, followed by a logical $X$ measurement on each of the 15 qubits in the Reed-Muller code. These measurement outcomes are used to reconstruct the four $X$-type stabilizers of the Reed-Muller code. If any stabilizer yields a measurement of $-1$, it indicates an error during the transversal $T$ gate, which is the noisiest part of the circuit. If no error is detected, the remaining qubit from the initial Bell pair is projected into a high-fidelity $\ket{T}$ state.}
    \label{fig:encoded_distillation}
\end{figure}

In the rest of this section, we will take as an example schemes for preparing $\ket{T}$ magic states on surface codes. Magic state distillation constructions are based on quantum codes that possess a transversal $T$ gate. The smallest distance-3 code with a transversal $T$ gate is the $[[15,1,3]]$ quantum Reed-Muller code, which uses 15 qubits to encode a single logical qubit~\cite{koutsioumpas2022smallest}. As illustrated in Figure~\ref{fig:Reed-Muller}, the code can be represented geometrically as a tetrahedron partitioned into four hexahedrons. Each hexahedron is associated with an X-type stabilizer supported on the qubits at its vertices, while the 18 $Z$ stabilizers are distributed over the hexahedron faces, 10 of which are linearly independent. Magic state distillation relies on code concatenation, where the Reed-Muller code serves as the outer code, and the surface codes as the inner codes. As depicted in Figure~\ref{fig:encoded_distillation}, the first step of the protocol is to prepare a logical Bell pair between two surface codes and encode one of them into the Reed-Muller code, using 14 additional logical qubits. Next, the transversal $T$ gate is applied to the Reed-Muller code by applying a $T^\dagger$ gate on each of the 15 surface codes. This is achieved by injecting a $\ket{T^\dagger} = \tfrac{1}{\sqrt{2}}(\ket{0} + e^{- i \pi/4} \ket{1})$ state onto an auxiliary surface code which is then consumed in the gate teleportation circuit shown in Figure~\ref{fig:T_teleport}. Each surface code within the Reed-Muller code is then measured in the $X$ basis, and the 4 $X$-type stabilizers of the Reed-Muller code are reconstructed and evaluated. As Clifford operations are performed with significantly higher fidelity than the $T^\dagger$ gates, the $X$-type stabilizers predominantly detect errors originating from the application of the latter. Moreover, since $X \ket{T^\dagger} = \sqrt{Z} \ket{T^\dagger}$, the $X$-type stabilizers are sufficient to detect errors in the $T^\dagger$ gates. If not all 4 $X$-type stabilizers yield a $+1$ outcome, the preparation is rejected, and the process is restarted. The remaining logical qubit from the initial Bell pair is now in the state $\ket{T}$. Since the Reed-Muller code is a distance-3 code, it can detect up to two $T^\dagger$ gate errors. If the input $\ket{T^\dagger}$ state error is $p$, the output error is $35p^3 + O(p^4)$, where the factor of 35 corresponds to the number of weight-3 undetected errors. As the preparation is rejected if any of the Reed-Muller $X$-type stabilizers flag an error, the acceptance probability of the scheme is given by $1-15p + O(p^2)$.

\begin{figure*}[t]
    \centering
    \includegraphics[width=0.8\textwidth]{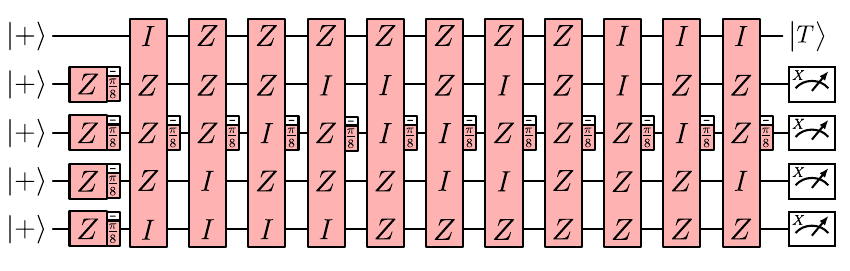}
    \caption{\textit{Unencoded} distillation protocol~\cite{haah2018codes,litinski2019magic}. This protocol uses 5 logical qubits and can be viewed as a space-time inversion of the circuit shown in Figure~\ref{fig:encoded_distillation}. Each gate corresponds to a single- or multi-qubit $-\pi/4$ $Z$ rotation of the form $e^{-i(-\frac{\pi}{8})Z \cdots Z}$, which can be implemented using a single $\ket{T}$ state. After performing 15 such logical rotations, the values of the four $X$-type stabilizers of the Reed-Muller code are evaluated from measurement results of logical qubits 2 to 5 in the $X$ basis. If all measurement outcomes are $+1$, the distillation is successful and the first logical qubit is in the state $\ket{T}$.}
    \label{fig:unencoded_distillation}
\end{figure*}

An alternative approach for constructing distillation circuits reduces the space cost at the expense of an increased time cost. It is commonly referred to as \textit{unencoded} distillation~\cite{haah2018codes,litinski2019magic}. Figure~\ref{fig:unencoded_distillation} shows such a distillation circuit for preparing a $\ket{T}$ magic state. The circuit consists of only 5 logical qubits instead of 15, with some of the $T^\dagger$ gates replaced by multi-Pauli $-\pi/4$ $Z$ rotations. The four $X$ measurements correspond to the four $X$ stabilizers of the Reed-Muller code, and the scheme is restarted if any of them yield a $-1$ outcome. Note that a single $\ket{T}$ state is needed for each multi-Pauli $-\pi/4$ $Z$ rotation, so that a total of 15 $\ket{T}$ states are required as inputs. In the surface code architecture, these rotations are performed via lattice surgery using ancillary qubit buses~\cite{horsman2012surface,fowler2018low,litinski2019game}. Since a bus can only be used for one rotation at a time, this limits the number of rotations that can be performed simultaneously. Using two buses and optimized scheduling reduces the total number of logical timesteps to six, with each timestep consisting of approximately $d$ syndrome extraction rounds, where $d$ represents the surface code distance~\cite{litinski2019magic}.

Since its introduction, the cost of distillation has been reduced. Notably, the distance of the inner surface codes that are measured in the $X$ basis does not need to be as high as that of the output surface code hosting the output $\ket{T}$ state~\cite{litinski2019magic}. For instance, in the unencoded scheme, the distance of logical qubits 2 to 5 can be reduced. Indeed, a $Z$ error on these logical qubits will commute with all rotations and be detected by the $X$ measurement, and thus rejected. Put differently, the outer Reed-Muller code can be used to correct both errors from the $-\pi/4$ rotation gates and topological errors from the surface codes. It has also been noted that any logical $X$ error on logical qubits 2 to 5 will be either trivial or detected by the $X$ measurements~\cite{lee2024low}.

\subsection{Clifford measurement}

Clifford measurement is an alternative to magic state distillation and has been shown to outperform it down to a logical error rate of $10^{-9}$ when the physical error rate is equal or lower than $10^{-3}$~\cite{gidney2024magic}. The scheme operates by measuring a logical Clifford operator that is transversal on the given code and stabilizes the target magic state~\cite{shor1996fault,gottesman1999demonstrating, zhou2000methodology}. The recent magic state cultivation construction proposes to measure $H_{XY} = (X+Y)/\sqrt{2}$ on the 2D color code to prepare a $\ket{T}$ state~\cite{zhou2000methodology,gidney2024magic}. For biased-noise qubits, it was also proposed to measure $X_1 (CX)_{23}$ on three repetition codes to prepare a $\ket{\text{CC}X} = \tfrac{1}{2}(\ket{000} + \ket{010} + \ket{100} + \ket{111})$ state~\cite{zhou2000methodology,chamberland2022building}. To ensure fault-tolerance, the Clifford measurement is performed with ancilla qubits prepared in a verified GHZ state, preventing error from propagating from the ancilla to the code. However, the GHZ state measurement itself is noisy and must be repeated and interleaved with postselected rounds of quantum error correction on the code.

Although this method has a very low qubit overhead (one surface code for the $H_{XY}$ measurement and three repetition codes for the $X_1 (CX)_{23}$ measurement),  the postselection process may incur a substantial time overhead. This results in a scheme that imposes stringent requirements on the physical error rate. For instance, for the $H_{XY}$ measurement, increasing the physical error rate from $p = 0.1\%$ to $p = 0.2\%$ leads to a two orders of magnitude increase in preparation time (see Figure 14 in Ref.~\cite{gidney2024magic}). Similarly, for the $X_1 (CX)_{23}$ measurement in the cat qubit architecture, increasing the physical phase-flip error rate from $p_Z = 10^{-3}$ to $p_Z = 5 \times 10^{-3}$ increases the logical error rate by almost three orders of magnitude, while also increasing the preparation time by two orders of magnitude (see Table~\ref{tab:overhead} entry \textit{'Bottom-up'}). Consequently, if biased-noise qubits are implemented with superconducting cat qubits, the Clifford measurement scheme imposes a very demanding requirement on the figure of merit of cat qubits $\kappa_1/\kappa_2$, with $\kappa_1$ representing the single-photon loss rate and $\kappa_2$ the two-photon dissipation rate used to stabilize cat qubits~\cite{guillaud2019repetition} (see Figure 15 in Ref.~\cite{chamberland2022building}).

\section{Main ideas}
\label{sec:main_ideas}

In this work, we focus on magic state distillation for biased-noise qubits and propose a scheme for efficiently implementing the $\pi/4$ logical rotation around the $X$ axis. In this section, we will assume for simplicity that qubits are infinitely biased and only suffer from phase-flip errors. The analysis of the case where the noise bias is finite will then be presented in Section~\ref{sec:surface_code}. One possible approach to distillation with infinitely biased-noise qubits is to adopt the same construction described in Figure~\ref{fig:encoded_distillation} or~\ref{fig:unencoded_distillation}, but to replace the surface code logical qubits with repetition codes. While substituting repetition codes reduces the space cost, the time cost remains essentially unchanged, since each logical operation still requires on the order of $\mathcal O(d)$ rounds, where $d$ is the distance of the repetition codes.

\begin{figure*}[ht]
    \centering
    \includegraphics[width=\textwidth]{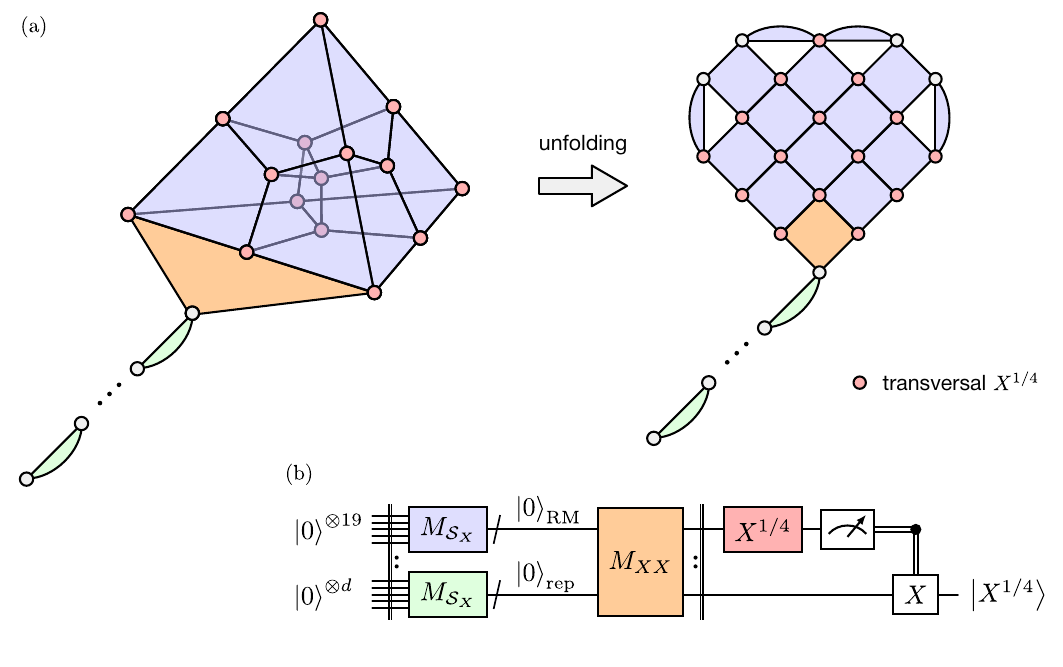}
    \caption{Unfolding of the $X$-type stabilizer group of the quantum Hadamard Reed-Muller code on a 2D layout. (a) Each qubit of the quantum Hadamard Reed-Muller code (left) corresponds to a red qubit in the \textit{unfolded} code (right). To enable nearest-neighbor connectivity, four additional data qubits are introduced in the unfolded code, along with weight-2 stabilizers. The repetition code that will host the output magic state (green) is merged with the unfolded code via a joint $X_\text{RM} X_\text{rep}$ measurement (orange). For practical implementation, it is convenient to add an extra data qubit and an ancilla qubit to facilitate the CNOT scheduling (see Appendix~\ref{app:Cnot ordering}). (b) Logical circuit corresponding to the unfolded distillation protocol. It begins by initializing all data qubits in the state $\ket{0}$, followed by the measurement of the $X$-type stabilizers of the unfolded code (blue), the $X_\text{RM} X_\text{rep}$ joint logical operator (orange) and the repetition code stabilizers (green) for several rounds. A transversal $X^{1/4}$ gate is then applied to the red qubits of the unfolded code, followed by the measurement of all data qubits of the unfolded code in the $Z$ basis. The four $Z$ stabilizers of the Hadamard Reed-Muller code are then reconstructed from these outcomes. If all stabilizers yield $+1$, the distillation is considered successful, and the repetition code is projected into the state $|X^{1/4}\rangle$. \href{https://algassert.com/crumble\#circuit=Q(0,4)0;Q(0,6)1;Q(1,3)2;Q(1,5)3;Q(1,7)4;Q(2,2)5;Q(2,4)6;Q(2,6)7;Q(2,8)8;Q(3,1)9;Q(3,3)10;Q(3,5)11;Q(3,7)12;Q(4,0)13;Q(4,2)14;Q(4,4)15;Q(4,6)16;Q(4,8)17;Q(5,1)18;Q(5,3)19;Q(5,5)20;Q(5,7)21;Q(6,0)22;Q(6,2)23;Q(6,4)24;Q(6,6)25;Q(6,8)26;Q(6,9)27;Q(6,10)28;Q(6,11)29;Q(6,12)30;Q(6,13)31;Q(6,14)32;Q(6,15)33;Q(6,16)34;Q(6,17)35;Q(6,18)36;Q(6,19)37;Q(6,20)38;Q(6,21)39;Q(6,22)40;Q(6,23)41;Q(6,24)42;Q(7,1)43;Q(7,3)44;Q(7,5)45;Q(7,7)46;Q(8,2)47;Q(8,4)48;Q(8,6)49;Q(8,8)50;Q(9,7)51;Q(10,8)52;R_13_22_5_14_23_47_0_6_15_24_48_1_7_16_25_49_8_17_26_50_52_28_30_32_34_36_38_40_42;RX_18_2_10_19_44_3_11_20_45_12_21_46_27_29_31_33_35_37_39_41;POLYGON(1,0,0,0.25)13_5;POLYGON(1,0,0,0.25)13_22_23_14;POLYGON(1,0,0,0.25)22_47;POLYGON(1,0,0,0.25)5_0;POLYGON(1,0,0,0.25)5_14_15_6;POLYGON(1,0,0,0.25)14_23_24_15;POLYGON(1,0,0,0.25)23_47_48_24;POLYGON(1,0,0,0.25)0_6_7_1;POLYGON(1,0,0,0.25)1_8;POLYGON(1,0,0,0.25)52_50;POLYGON(1,0,0,0.25)6_15_16_7;POLYGON(1,0,0,0.25)15_24_25_16;POLYGON(1,0,0,0.25)24_48_49_25;POLYGON(1,0,0,0.25)7_16_17_8;POLYGON(1,0,0,0.25)16_25_26_17;POLYGON(1,0,0,0.25)25_49_50_26;POLYGON(1,0,0,0.25)26_28;POLYGON(1,0,0,0.25)28_30;POLYGON(1,0,0,0.25)30_32;POLYGON(1,0,0,0.25)32_34;POLYGON(1,0,0,0.25)34_36;POLYGON(1,0,0,0.25)36_38;POLYGON(1,0,0,0.25)38_40;POLYGON(1,0,0,0.25)40_42;TICK;CX_18_13_2_5_10_15_19_23_44_48_3_6_11_7_20_24_45_25_12_8_21_16_46_26_27_28_29_30_31_32_33_34_35_36_37_38_39_40_41_42;TICK;CX_18_22_2_0_10_5_19_14_44_47_3_1_11_15_20_25_45_24_12_16_21_17_46_50_27_26_29_28_31_30_33_32_35_34_37_36_39_38_41_40;RX_9_43_4_51;TICK;CX_9_5_18_23_43_47_10_14_19_15_44_24_3_0_11_6_20_16_45_49_4_8_12_7_21_26_46_25_51_52;MX_2_27_29_31_33_35_37_39_41;TICK;CX_9_13_18_14_43_22_10_6_19_24_44_23_3_7_11_16_20_15_45_48_4_1_12_17_21_25_46_49_51_50;RX_27_29_31_33_35_37_39_41;TICK;CX_27_26_29_28_31_30_33_32_35_34_37_36_39_38_41_40;MX_9_18_43_10_19_44_3_11_20_45_4_12_21_46_51;TICK;CX_27_28_29_30_31_32_33_34_35_36_37_38_39_40_41_42;RX_18_2_10_19_44_3_11_20_45_12_21_46;TICK;CX_18_13_2_5_10_15_19_23_44_48_3_6_11_7_20_24_45_25_12_8_21_16_46_26;MX_27_29_31_33_35_37_39_41;DT(6,9,0)rec[-8]_rec[-31];DT(6,11,0)rec[-7]_rec[-30];DT(6,13,0)rec[-6]_rec[-29];DT(6,15,0)rec[-5]_rec[-28];DT(6,17,0)rec[-4]_rec[-27];DT(6,19,0)rec[-3]_rec[-26];DT(6,21,0)rec[-2]_rec[-25];DT(6,23,0)rec[-1]_rec[-24];TICK;CX_18_22_2_0_10_5_19_14_44_47_3_1_11_15_20_25_45_24_12_16_21_17_46_50;RX_9_43_4_51_27_29_31_33_35_37_39_41;TICK;CX_9_5_18_23_43_47_10_14_19_15_44_24_3_0_11_6_20_16_45_49_4_8_12_7_21_26_46_25_51_52_27_28_29_30_31_32_33_34_35_36_37_38_39_40_41_42;MX_2;DT(1,3,1)rec[-1]_rec[-33];TICK;CX_9_13_18_14_43_22_10_6_19_24_44_23_3_7_11_16_20_15_45_48_4_1_12_17_21_25_46_49_51_50_27_26_29_28_31_30_33_32_35_34_37_36_39_38_41_40;TICK;MX_9_18_43_10_19_44_3_11_20_45_4_12_21_46_51_27_29_31_33_35_37_39_41;DT(3,1,2)rec[-23]_rec[-47];DT(5,1,2)rec[-22]_rec[-46];DT(7,1,2)rec[-21]_rec[-45];DT(3,3,2)rec[-20]_rec[-44];DT(5,3,2)rec[-19]_rec[-43];DT(7,3,2)rec[-18]_rec[-42];DT(1,5,2)rec[-17]_rec[-41];DT(3,5,2)rec[-16]_rec[-40];DT(5,5,2)rec[-15]_rec[-39];DT(7,5,2)rec[-14]_rec[-38];DT(1,7,2)rec[-13]_rec[-37];DT(3,7,2)rec[-12]_rec[-36];DT(5,7,2)rec[-11]_rec[-35];DT(7,7,2)rec[-10]_rec[-34];DT(9,7,2)rec[-9]_rec[-33];DT(6,9,2)rec[-8]_rec[-32];DT(6,11,2)rec[-7]_rec[-31];DT(6,13,2)rec[-6]_rec[-30];DT(6,15,2)rec[-5]_rec[-29];DT(6,17,2)rec[-4]_rec[-28];DT(6,19,2)rec[-3]_rec[-27];DT(6,21,2)rec[-2]_rec[-26];DT(6,23,2)rec[-1]_rec[-25];TICK;RX_18_2_10_19_44_3_11_20_45_12_21_46_27_29_31_33_35_37_39_41;TICK;CX_18_13_2_5_10_15_19_23_44_48_3_6_11_7_20_24_45_25_12_8_21_16_46_26_27_28_29_30_31_32_33_34_35_36_37_38_39_40_41_42;TICK;CX_18_22_2_0_10_5_19_14_44_47_3_1_11_15_20_25_45_24_12_16_21_17_46_50_27_26_29_28_31_30_33_32_35_34_37_36_39_38_41_40;RX_9_43_4_51;TICK;CX_9_5_18_23_43_47_10_14_19_15_44_24_3_0_11_6_20_16_45_49_4_8_12_7_21_26_46_25_51_52;MX_2_27_29_31_33_35_37_39_41;DT(1,3,3)rec[-9]_rec[-33];DT(6,9,3)rec[-8]_rec[-17];DT(6,11,3)rec[-7]_rec[-16];DT(6,13,3)rec[-6]_rec[-15];DT(6,15,3)rec[-5]_rec[-14];DT(6,17,3)rec[-4]_rec[-13];DT(6,19,3)rec[-3]_rec[-12];DT(6,21,3)rec[-2]_rec[-11];DT(6,23,3)rec[-1]_rec[-10];TICK;CX_9_13_18_14_43_22_10_6_19_24_44_23_3_7_11_16_20_15_45_48_4_1_12_17_21_25_46_49_51_50;RX_27_29_31_33_35_37_39_41;TICK;CX_27_26_29_28_31_30_33_32_35_34_37_36_39_38_41_40;MX_9_18_43_10_19_44_3_11_20_45_4_12_21_46_51;DT(3,1,4)rec[-15]_rec[-47];DT(5,1,4)rec[-14]_rec[-46];DT(7,1,4)rec[-13]_rec[-45];DT(3,3,4)rec[-12]_rec[-44];DT(5,3,4)rec[-11]_rec[-43];DT(7,3,4)rec[-10]_rec[-42];DT(1,5,4)rec[-9]_rec[-41];DT(3,5,4)rec[-8]_rec[-40];DT(5,5,4)rec[-7]_rec[-39];DT(7,5,4)rec[-6]_rec[-38];DT(1,7,4)rec[-5]_rec[-37];DT(3,7,4)rec[-4]_rec[-36];DT(5,7,4)rec[-3]_rec[-35];DT(7,7,4)rec[-2]_rec[-34];DT(9,7,4)rec[-1]_rec[-33];TICK;CX_27_28_29_30_31_32_33_34_35_36_37_38_39_40_41_42;RX_18_2_10_19_44_3_11_20_45_12_21_46;TICK;CX_18_13_2_5_10_15_19_23_44_48_3_6_11_7_20_24_45_25_12_8_21_16_46_26;MX_27_29_31_33_35_37_39_41;DT(6,9,5)rec[-8]_rec[-31];DT(6,11,5)rec[-7]_rec[-30];DT(6,13,5)rec[-6]_rec[-29];DT(6,15,5)rec[-5]_rec[-28];DT(6,17,5)rec[-4]_rec[-27];DT(6,19,5)rec[-3]_rec[-26];DT(6,21,5)rec[-2]_rec[-25];DT(6,23,5)rec[-1]_rec[-24];TICK;CX_18_22_2_0_10_5_19_14_44_47_3_1_11_15_20_25_45_24_12_16_21_17_46_50;RX_9_43_4_51_27_29_31_33_35_37_39_41;TICK;CX_9_5_18_23_43_47_10_14_19_15_44_24_3_0_11_6_20_16_45_49_4_8_12_7_21_26_46_25_51_52_27_28_29_30_31_32_33_34_35_36_37_38_39_40_41_42;MX_2;DT(1,3,6)rec[-1]_rec[-33];TICK;CX_9_13_18_14_43_22_10_6_19_24_44_23_3_7_11_16_20_15_45_48_4_1_12_17_21_25_46_49_51_50_27_26_29_28_31_30_33_32_35_34_37_36_39_38_41_40;TICK;MX_9_18_43_10_19_44_3_11_20_45_4_12_21_46_51_27_29_31_33_35_37_39_41;DT(3,1,7)rec[-23]_rec[-47];DT(5,1,7)rec[-22]_rec[-46];DT(7,1,7)rec[-21]_rec[-45];DT(3,3,7)rec[-20]_rec[-44];DT(5,3,7)rec[-19]_rec[-43];DT(7,3,7)rec[-18]_rec[-42];DT(1,5,7)rec[-17]_rec[-41];DT(3,5,7)rec[-16]_rec[-40];DT(5,5,7)rec[-15]_rec[-39];DT(7,5,7)rec[-14]_rec[-38];DT(1,7,7)rec[-13]_rec[-37];DT(3,7,7)rec[-12]_rec[-36];DT(5,7,7)rec[-11]_rec[-35];DT(7,7,7)rec[-10]_rec[-34];DT(9,7,7)rec[-9]_rec[-33];DT(6,9,7)rec[-8]_rec[-32];DT(6,11,7)rec[-7]_rec[-31];DT(6,13,7)rec[-6]_rec[-30];DT(6,15,7)rec[-5]_rec[-29];DT(6,17,7)rec[-4]_rec[-28];DT(6,19,7)rec[-3]_rec[-27];DT(6,21,7)rec[-2]_rec[-26];DT(6,23,7)rec[-1]_rec[-25];TICK;RX_18_2_10_19_44_3_11_20_45_12_21_46_27_29_31_33_35_37_39_41;TICK;CX_18_13_2_5_10_15_19_23_44_48_3_6_11_7_20_24_45_25_12_8_21_16_46_26_27_28_29_30_31_32_33_34_35_36_37_38_39_40_41_42;TICK;CX_18_22_2_0_10_5_19_14_44_47_3_1_11_15_20_25_45_24_12_16_21_17_46_50_27_26_29_28_31_30_33_32_35_34_37_36_39_38_41_40;RX_9_43_4_51;TICK;CX_9_5_18_23_43_47_10_14_19_15_44_24_3_0_11_6_20_16_45_49_4_8_12_7_21_26_46_25_51_52;MX_2_27_29_31_33_35_37_39_41;DT(1,3,8)rec[-9]_rec[-33];DT(6,9,8)rec[-8]_rec[-17];DT(6,11,8)rec[-7]_rec[-16];DT(6,13,8)rec[-6]_rec[-15];DT(6,15,8)rec[-5]_rec[-14];DT(6,17,8)rec[-4]_rec[-13];DT(6,19,8)rec[-3]_rec[-12];DT(6,21,8)rec[-2]_rec[-11];DT(6,23,8)rec[-1]_rec[-10];TICK;CX_9_13_18_14_43_22_10_6_19_24_44_23_3_7_11_16_20_15_45_48_4_1_12_17_21_25_46_49_51_50;RX_27_29_31_33_35_37_39_41;TICK;CX_27_26_29_28_31_30_33_32_35_34_37_36_39_38_41_40;MX_9_18_43_10_19_44_3_11_20_45_4_12_21_46_51;DT(3,1,9)rec[-15]_rec[-47];DT(5,1,9)rec[-14]_rec[-46];DT(7,1,9)rec[-13]_rec[-45];DT(3,3,9)rec[-12]_rec[-44];DT(5,3,9)rec[-11]_rec[-43];DT(7,3,9)rec[-10]_rec[-42];DT(1,5,9)rec[-9]_rec[-41];DT(3,5,9)rec[-8]_rec[-40];DT(5,5,9)rec[-7]_rec[-39];DT(7,5,9)rec[-6]_rec[-38];DT(1,7,9)rec[-5]_rec[-37];DT(3,7,9)rec[-4]_rec[-36];DT(5,7,9)rec[-3]_rec[-35];DT(7,7,9)rec[-2]_rec[-34];DT(9,7,9)rec[-1]_rec[-33];TICK;CX_27_28_29_30_31_32_33_34_35_36_37_38_39_40_41_42;SQRT_X_24_7_16_25_50_13_48_14_15_6_23_22_17_49_1;TICK;M_13_22_5_14_23_47_0_6_15_24_48_1_7_16_25_49_8_17_50_52;MX_27_29_31_33_35_37_39_41;DT(6,9,10)rec[-8]_rec[-51];DT(6,11,10)rec[-7]_rec[-50];DT(6,13,10)rec[-6]_rec[-49];DT(6,15,10)rec[-5]_rec[-48];DT(6,17,10)rec[-4]_rec[-47];DT(6,19,10)rec[-3]_rec[-46];DT(6,21,10)rec[-2]_rec[-45];DT(6,23,10)rec[-1]_rec[-44];OI(0)rec[-18]_rec[-19]_rec[-20]_rec[-21]_rec[-22]_rec[-23]_rec[-24]_rec[-25]_rec[-26]_rec[-27]_rec[-28]_rec[-150]_rec[-152]_rec[-153]_rec[-155];OI(1)rec[-9]_rec[-10]_rec[-12]_rec[-13]_rec[-16]_rec[-17]_rec[-18]_rec[-21]_rec[-22]_rec[-23]_rec[-26]_rec[-27]_rec[-28]_rec[-142]_rec[-143]_rec[-144]_rec[-145]_rec[-150]_rec[-151]_rec[-152]_rec[-153]_rec[-155];OI(2)rec[-9]_rec[-10]_rec[-11]_rec[-13]_rec[-15]_rec[-18]_rec[-20]_rec[-23]_rec[-25]_rec[-27]_rec[-142]_rec[-143]_rec[-150]_rec[-151]_rec[-153];OI(3)rec[-11]_rec[-12]_rec[-13]_rec[-14]_rec[-17]_rec[-20]_rec[-21]_rec[-23]_rec[-24]_rec[-27]_rec[-144]_rec[-145]_rec[-146]_rec[-148]_rec[-150]_rec[-153];TICK;S_26_28_30_32_34_36_38_40_42;TICK;CZ_26_28;TICK;CZ_26_30;TICK;CZ_26_32;TICK;CZ_26_34;TICK;CZ_26_36;TICK;CZ_26_38;TICK;CZ_26_40;TICK;CZ_26_42_28_30;TICK;CZ_28_32;TICK;CZ_28_34;TICK;CZ_28_36;TICK;CZ_28_38;TICK;CZ_28_40;TICK;CZ_28_42_30_32;TICK;CZ_30_34;TICK;CZ_30_36;TICK;CZ_30_38;TICK;CZ_30_40;TICK;CZ_30_42_32_34;TICK;CZ_32_36;TICK;CZ_32_38;TICK;CZ_32_40;TICK;CZ_32_42_34_36;TICK;CZ_34_38;TICK;CZ_34_40;TICK;CZ_34_42_36_38;TICK;CZ_36_40;TICK;CZ_36_42_38_40;TICK;CZ_38_42;TICK;CZ_40_42;TICK;MX_26_28_30_32_34_36_38_40_42;DT(6,9,11)rec[-8]_rec[-9]_rec[-17];DT(6,11,11)rec[-7]_rec[-8]_rec[-16];DT(6,13,11)rec[-6]_rec[-7]_rec[-15];DT(6,15,11)rec[-5]_rec[-6]_rec[-14];DT(6,17,11)rec[-4]_rec[-5]_rec[-13];DT(6,19,11)rec[-3]_rec[-4]_rec[-12];DT(6,21,11)rec[-2]_rec[-3]_rec[-11];DT(6,23,11)rec[-1]_rec[-2]_rec[-10];OI(4)rec[-9]_rec[-18]_rec[-19]_rec[-20]_rec[-21]_rec[-26]_rec[-31]_rec[-32]_rec[-33]_rec[-34]_rec[-35]_rec[-36]_rec[-37]_rec[-151]_rec[-153]_rec[-154]_rec[-155]_rec[-157]_rec[-159]_rec[-161]_rec[-162]_rec[-164]}{The circuit can be found in Crumble.}}
    \label{fig:unfolding}
\end{figure*}

Here, we propose an alternative and extremely economical construction that replaces each surface code of the distillation protocol with a single physical biased-noise qubit while preserving the same fidelity for the resulting magic state. Usually, distillation protocols are performed at the logical level so that Clifford gate errors are negligible relative to those of the input states $\ket{T}$. Indeed, if distillation is instead performed at the physical level, for example with the circuit of Figure~\ref{fig:encoded_distillation}, fewer than three errors could remain undetected. This occurs because the encoding circuit is not fault-tolerant and propagates errors. A fault-tolerant alternative would be to prepare the Reed-Muller codespace by measuring the stabilizers rather than using an encoding circuit. However, because the Reed-Muller code is three-dimensional, this method is not compatible with a two-dimensional qubit architecture.

In contrast, if qubits only experience phase-flip errors, only the measurement of $X$-type stabilizers is necessary. Indeed, the Reed-Muller codespace can be prepared fault-tolerantly by first initializing all 15 physical qubits in $\ket{0}$ and then measuring the $X$-type stabilizers. However, the $X$-type stabilizers of the Reed-Muller code are weight-8 stabilizers, which are difficult to measure at the physical level. Consequently, it is convenient to swap the $X$ and $Z$ bases and consider the Hadamard Reed-Muller code. For the remainder of this work, the $X$-type stabilizers will denote the weight-4 stabilizers (the code faces, see Figure~\ref{fig:Reed-Muller}), and $Z$-type stabilizers will denote the weight-8 stabilizers (the code volumes). Note that the Hadamard Reed-Muller code features a transversal $X^{1/4}$ gate, so the protocol now distills $\ket{X^{1/4}} = \text{cos}(\pi/8)\ket{0} - i\ \text{sin}(\pi/8)\ket{1}$ state, which is also a magic state. One might be concerned that the $X^{1/4}$ gate is not bias-preserving, but this is not an issue, as any induced bit-flip is detected by the $Z$-type stabilizers of the Hadamard Reed-Muller code measured after the application of the transversal gate.

Our main result is that the $X$-type stabilizers of the Hadamard Reed-Muller code can be measured simply with a two-dimensional qubit layout. As mentioned previously, this enables the fault-tolerant preparation of the Hadamard Reed-Muller codespace at the physical level in 2D, allowing for distillation at this level as well. Figure~\ref{fig:unfolding} illustrates the unfolding of the $X$-type stabilizer group of the Hadamard Reed-Muller code onto a two-dimensional nearest-neighbor grid. In this new \textit{unfolded} code, some weight-2 $X$ stabilizers are introduced in addition to weight-4 stabilizers, and the total number of data qubits increases from 15 to 19. These additional checks and qubits serve as buses, enabling a strictly nearest-neighbor layout. Consequently, the $X$-stabilizer group of the Hadamard Reed-Muller code is a subgroup of the $X$-stabilizer group of the unfolded code. We note that, in the standard Reed-Muller code, the required weight-8 volume stabilizers would not only need to be measured but also cannot be unfolded into a two-dimensional layout.

As in standard distillation protocols, a Bell pair must be created between the Hadamard Reed-Muller code and the logical output qubit that will host the magic state. This can be achieved by merging a repetition code with the unfolded code through the measurement of the (orange) $X$-type stabilizer as shown in Figure~\ref{fig:unfolding}. Overall, the protocol begins by preparing all data qubits of both the unfolded code and the repetition code in the state $\ket{0}$, then measuring the $X$-type stabilizers including the merging stabilizer over several rounds. Next, the transversal $X^{1/4}$ gate is applied to the unfolded code by applying physical $X^{-1/4}$ gates to the subset of qubits belonging to the Hadamard Reed-Muller subgroup. The qubits of the unfolded code are then measured in the $Z$ basis, thereby enabling the reconstruction of the $Z$-type stabilizers of the Hadamard Reed-Muller code and the detection of any faulty $X^{-1/4}$ rotations. If the product of all $Z$-basis measurement outcomes is $-1$, an $X$ feedback operation is applied to the repetition code, which is then in the state $\ket{X^{1/4}}$.

\section{Distillation with the unfolded code}
\label{sec:repetition code}

In this section, we provide a detailed description of the fault-tolerant features of the magic state preparation protocol and the construction of the unfolded code.

\subsection{Results}

\begin{figure}[t]
    \centering
    \includegraphics[width=0.48\textwidth]{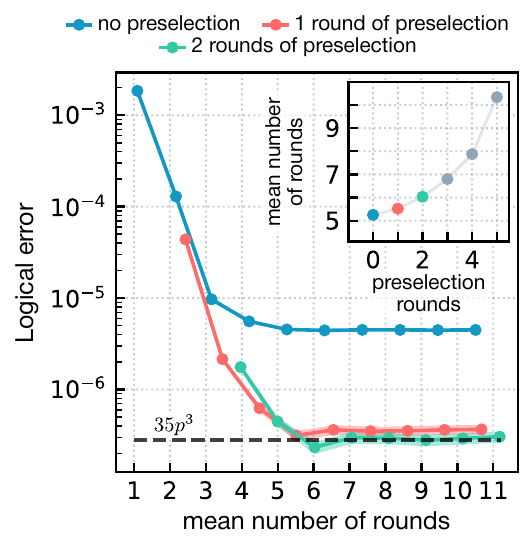}
    \caption{Logical error probability of the $|X^{1/4}\rangle$ magic state prepared using the unfolded distillation protocol. A circuit-level physical phase-flip error rate of $p_Z = 10^{-3}$ per operation is assumed. Since $X$-type stabilizer measurements are noisy, a minimum number of rounds is required to reach the expected distillation logical error rate of $35p^3$, where $p$ is the error probability of the $X^{-1/4}$ gate. \textit{Preselection} refers to repeatedly measuring the $X$-type stabilizers until, for a specified number of consecutive rounds, all outcomes are constant (\textit{i.e.}, no error reported). For example, for 2 rounds of preselection (curve shown in green), at least three rounds of stabilizer measurement are performed. This includes an initial round that randomly projects the $X$-type stabilizer, followed by at least two additional rounds to allow for preselection. Each subsequent point on the curve adds one round to the fixed number of initial rounds. The mean number of rounds then accounts for the extra time required when preselection fails, as well as the final acceptance check based on the $Z$-type stabilizers of the Hadamard Reed-Muller code. The numerical results show that one round of preselection and approximately 5.5 total rounds are necessary to reach the maximum achievable magic state fidelity. The inset displays the mean number of rounds as a function of the number of preselection rounds for a fixed number of initial rounds of 5. The scaling is exponential, indicating that preselection becomes impractical for higher physical error rates.}
    \label{fig:repetition_code}
\end{figure}

As described in Section~\ref{sec:main_ideas}, the first part of the protocol involves measuring the $X$ stabilizers following the initialization of the qubits in the $Z$ basis, to prepare the logical $\ket{0}$ of the Hadamard Reed-Muller code. But if only one round of $X$ stabilizer measurements is performed, some measurement errors could lead to a logical error in the prepared magic state. Thus, several rounds of stabilizer measurements must be performed before applying the transversal $X^{1/4}$ gate. Note that this is not surprising, as the unfolded code loses the single-shot property of the 3D Reed-Muller code. Indeed, the redundancy in the stabilizer measurements, which is the origin of this single-shot property, is removed by the unfolding. Figure~\ref{fig:repetition_code} shows (in blue) the fidelity of the output magic state as a function of the number of stabilizer measurement rounds. As expected, the logical error decreases exponentially until it reaches a plateau, where the number of rounds is sufficient to prepare the Hadamard Reed-Muller codespace with sufficiently high fidelity. The dotted line represents the expected logical error rate when the distillation is limited solely by the fidelity of the $X^{-1/4}$ gates, which is given by $35p^3$ where $p$ is the error probability of the $X^{-1/4}$ gate. However, a gap of more than one order of magnitude can be observed when compared to the expected logical error rate. This is explained by the fact that measurement errors on the final $X$ stabilizer measurements are equivalent to residual $Z$ errors on the data qubits of the unfolded code. These $Z$ errors are then mapped to $X^{-1/2} Z$ after the $X^{-1/4}$ gates, eventually flipping (with 50\% probability) the outcome of the $Z$ stabilizer measurements of the Hadamard Reed-Muller code. These error configurations effectively increase the $X^{-1/4}$ gate input probability $p$, thereby reducing the logical fidelity of the output magic state.

A solution proposed in Ref.~\cite{litinski2019magic} and fairly similar to the STOP algorithm proposed in Ref.~\cite{chamberland2022building} consists in \textit{preselecting} on the last round(s) of stabilizer measurements. That is, the transversal $X^{1/4}$ is applied only if, for a certain number of prior rounds, all syndrome measurements are constant, meaning no error was reported during these rounds. If this is not the case after the required number of rounds necessary for the preparation of the Hadamard Reed-Muller codespace, the transversal $X^{1/4}$ gate is delayed until the condition is met. Figure~\ref{fig:repetition_code} shows the logical error of the output magic state for different numbers of preselected rounds. For a circuit-level phase-flip noise with error probability $p_Z=10^{-3}$ per operation, 5 rounds including 1 to 2 preselected are necessary for the distillation to be limited solely by the $X^{-1/4}$ gate errors. In practice, for a physical error rate of $p_Z = 10^{-3}$, the total number of rounds is only slightly increased due to preselection. Including as well the usual distillation postselection on the $Z$-type Hadamard Reed-Muller stabilizers, this results in an average total number of rounds smaller than 6, as can be seen in the inset of Figure~\ref{fig:repetition_code}. Moreover, the distance of the repetition code must be chosen such that it does not limit the overall fidelity of the magic state. Here, a distance-9 repetition code is sufficient to meet this requirement.

We note that the presented scheme requires decoding with a reaction time on the order of the duration of a stabilizer measurement cycle. Indeed, if a $Z$ error occurs on a data qubit, its Pauli frame must be updated, effectively transforming the $X^{-1/4}$ rotation into an $X^{1/4}$ rotation. In practice, this implies that the decoding of $X$-type stabilizers must be completed prior to applying the $X^{\pm 1/4}$ rotations. If the classical hardware and decoding technique do not meet these requirements, we present an alternative scheme in Appendix~\ref{app:Clifford ladder} which alleviates the need for fast decoding at the cost of approximately doubling the number of cycles. This adapted scheme, because it also removes the need for preselection, is beneficial as well in situations where the physical error rate is high, and the average number of rounds required due to preselection becomes excessive.\pagebreak

Overall, the logical error rate of the scheme presented in this section is given by $35(p^{\text{CNOT, target}}_Z + p^{\text{idle, mmt}}_{Z} + p^{X^{-1/4}}_X + p^{X^{-1/4}}_Y)^3$, where $p^{\text{CNOT, target}}_Z$ denotes the probability of a $Z$ error on the target (data) qubits of the CNOT gates prior to the $X^{-1/4}$ gates. These CNOT operations are used in the $X$-stabilizer measurements and only the target $Z$ errors of the final measurement round have an impact on the distillation error budget. Next, $p^{\text{idle, mmt}}_{Z}$ denotes the probability of a $Z$ error during the $X$ measurement of the final round of X-stabilizer measurements. Indeed, given that preselection is used, it is crucial to check the outcome of the final round of $X$ stabilizer measurements before deciding whether to apply the transversal $X^{1/4}$ gate or to continue the stabilizer measurements. Finally, $p^{X^{-1/4}}_X$ and $p^{X^{-1/4}}_Y$ are the probabilities of $X$ and $Y$ errors, respectively, on the $X^{-1/4}$ gate. If the scheme of Appendix~\ref{app:Clifford ladder} is used, the logical error is lower bounded by $35(p^{\text{CNOT, target}}_Z + p^{X^{-1/4}}_X + p^{X^{-1/4}}_Y)^3$ as the $X^{\pm 1/4}$ gate can be applied directly after the CNOT gate (see Appendix~\ref{app:Clifford ladder} for more details). This lower bound is achieved when $p_Z = 10^{-3}$, but the logical error rate increases when $p_Z = 5 \times 10^{-3}$ due to greater contributions from certain weight-4 error configurations (see Appendix~\ref{app:high error} for more details). Nevertheless, the scheme can remain more effective than the one presented in this section (with preselection), even with the increase in logical error rate. We give more details about the decoding and simulation techniques in Appendix~\ref{app:simulation}.

\subsection{Code construction}

\begin{figure*}[t]
    \centering
    \includegraphics[width=\textwidth]{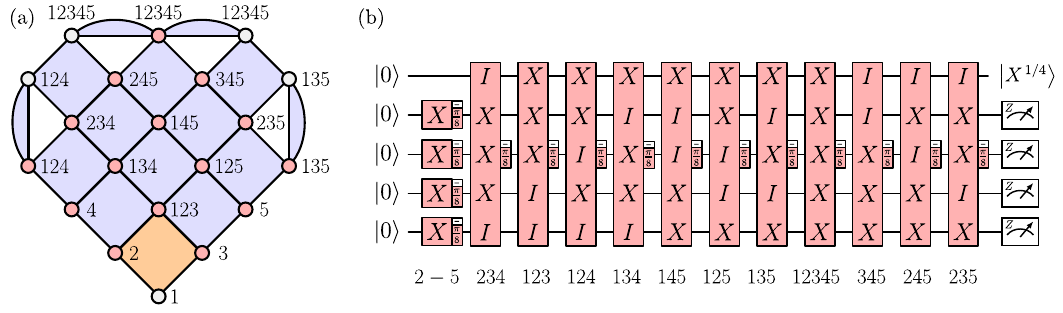}
    \caption{Unfolded code construction. The unfolded code shown in (a) can be interpreted as an unfolding of the quantum Hadamard Reed-Muller code, but it can also be viewed as a phase-flip code in which each $Z$-type stabilizer of the Hadamard Reed-Muller code is promoted to a logical qubit, resulting in a total of 5 logical qubits. The number attached to each qubit represents the support of the $Z$ logical operators 1 to 5. The unencoded distillation circuit shown in (b) then corresponds to the logical operations performed on this unfolded code when treated as a phase-flip code during the transversal application of the $X^{1/4}$ gate. Each $X^{-1/4}$ gate applied to a physical qubit in the unfolded code (a) implements, at the logical level, one of the 15 single- or multi-qubit $X$-type rotations from the unencoded distillation circuit (b). Therefore, an unfolded code can be constructed by designing a code with 5 logical qubits such that each logical operator in the set $\mathcal S _{RM} = \{ X_{L,i} \mid 1 \le i \le 5 \} \cup \{ X_{L,i} X_{L,j} X_{L,k} \mid 1 \le i, j, k \le 5, i \ne j \ne k \} \cup \{ X_{L,1} X_{L,2} X_{L,3} X_{L,4} X_{L,5} \}$ corresponds to a physical $X$ operator (see Appendix~\ref{app:code search} for more details).}
    \label{fig:mapping XL}
\end{figure*}

We next describe the construction of the unfolded code. Although it is convenient to view the construction as an unfolding of the 3D Hadamard Reed-Muller code, finding a 2D unfolding appears to be a difficult problem. Nevertheless, an equivalent formulation of the problem allows for a more constructive approach. The unfolded code, containing only $X$-type stabilizers, can be interpreted as a 2D phase-flip code that admits a transversal $X^{1/4}$ gate. In this perspective, alongside the logical qubit encoded by the Hadamard Reed-Muller code, the four $Z$-type stabilizers of the code are reinterpreted as logical qubits, resulting in a phase-flip code encoding a total of five logical qubits. While the $Z$ logical operators have support on multiple physical qubits, the code does not protect against bit-flip errors, and as a result, each $X$ logical operator has support on a single physical qubit. Thus, an $X^{-1/4}$ gate applied to one of these physical qubits corresponds to a logical $X^{-1/4}_L$ gate on the associated logical qubit. However, some physical qubits are in the support of multiple $Z$ logical operators. Applying a physical $X^{-1/4}$ gate to a qubit lying in the support of $Z_{L,i_1}, \ldots, Z_{L,i_n}$ corresponds to applying the logical gate $\exp\left(-\frac{i}{2} \frac{\pi}{4} X_{L,i_1} \cdots X_{L,i_n}\right)$. We note that, in a different context, the LHZ architecture employs similar techniques to compile algorithms by enabling all-to-all connectivity at the logical level, despite having only nearest-neighbor connectivity at the physical level~\cite{lechner2015quantum}.

Figure~\ref{fig:mapping XL}(a) shows the support of the five $Z$ logical operators of the unfolded code, or equivalently, the logical $X$ gate corresponding to a physical $X$ gate applied to each qubit. Overall, in the phase-flip code picture, the full set of 15 $X^{-1/4}$ gates corresponds, at the logical level, to the gate sequence of the unencoded distillation circuit shown in Figure~\ref{fig:unencoded_distillation} (in the Hadamard-conjugated basis), as well as in Figure~\ref{fig:mapping XL}(b). Equivalently, the unfolded code permits the entire block of 15 logical gates on the 5 logical qubits to be executed transversally. Thus, unfolding the Hadamard Reed-Muller code amounts to finding a 2D planar phase-flip code in which each of the 15 logical operations shown in Figure~\ref{fig:mapping XL}(b) is implemented by a single-qubit $X^{-1/4}$ gate.

Consequently, the code can be built iteratively, beginning with five physical qubits that, by definition, constitute the support of the logical operators $X_{L,1}$-$X_{L,5}$. In the next step, we could add, for example, a stabilizer that acts on the three qubits supporting $X_{L,1}$, $X_{L,2}$, and $X_{L,3}$ together with a newly added physical qubit. This new qubit then becomes the support of the logical operator $X_{L,123}$. The objective is to select physical qubits and stabilizers such that, for each of the 15 logical rotations depicted in Figure~\ref{fig:mapping XL}(b), there is at least one physical qubit on which a single-qubit $X^{-1/4}$ gate implements that logical rotation. This formulation allows us to reduce the task to a Boolean satisfiability (SAT) instance, as detailed in Appendix~\ref{app:code search}.

\section{Scheme with moderate noise bias}
\label{sec:surface_code}

\begin{figure*}[t]
    \centering
    \includegraphics[width=\textwidth]{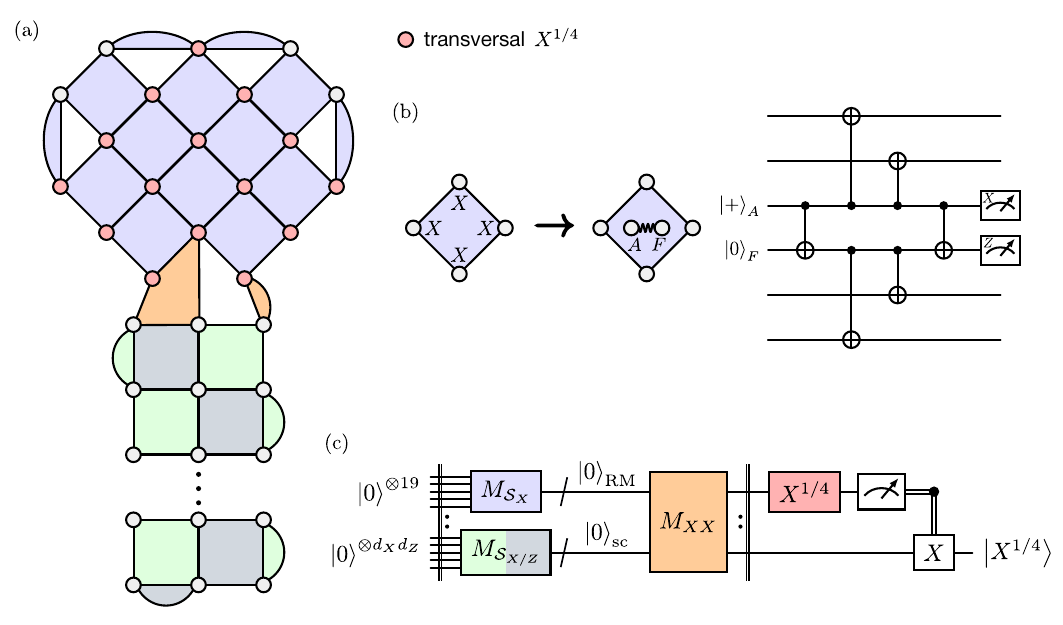}
    \caption{Unfolded distillation in a thin surface code for magic state preparation at modest noise bias. The $Z$-type stabilizers of the Hadamard Reed-Muller code, reconstructed at the end of the unfolded distillation protocol, not only detect errors on the $X^{-1/4}$ gates, but also detect bit-flip errors on the data qubits of the unfolded code. As a result, the unfolded code is resilient to up to two bit-flip errors and, when merged with a thin surface code, can prepare magic states under moderate noise bias, as shown in (a). However, bit-flip errors on the ancilla qubits used in the unfolded code can propagate to two data qubits, compromising the bit-flip protection. To mitigate this, stabilizer measurements can be performed using a flag qubit, as illustrated in (b). Panel (c) shows the logical circuit corresponding to this protocol. \href{https://algassert.com/crumble\#circuit=Q(0,4)0;Q(0,6)1;Q(1,3)2;Q(1,5)3;Q(1,7)4;Q(2,2)5;Q(2,4)6;Q(2,5)7;Q(2,6)8;Q(2,8)9;Q(3,1)10;Q(3,3)11;Q(3,5)12;Q(3,7)13;Q(4,0)14;Q(4,2)15;Q(4,3)16;Q(4,4)17;Q(4,5)18;Q(4,6)19;Q(4,7)20;Q(4,8)21;Q(5,1)22;Q(5,3)23;Q(5,5)24;Q(5,7)25;Q(6,0)26;Q(6,1)27;Q(6,2)28;Q(6,3)29;Q(6,4)30;Q(6,5)31;Q(6,6)32;Q(6,7)33;Q(6,8)34;Q(6,9)35;Q(7,1)36;Q(7,3)37;Q(7,5)38;Q(7,7)39;Q(7,8)40;Q(7,9)41;Q(8,2)42;Q(8,3)43;Q(8,4)44;Q(8,5)45;Q(8,6)46;Q(8,7)47;Q(8,8)48;Q(8,9)49;Q(8,10)50;Q(8,11)51;Q(9,7)52;Q(9,8)53;Q(9,9)54;Q(9,10)55;Q(9,11)56;Q(10,7)57;Q(10,8)58;Q(10,9)59;Q(10,10)60;Q(10,11)61;Q(10,12)62;Q(10,13)63;Q(11,9)64;Q(11,10)65;Q(11,11)66;Q(11,12)67;Q(11,13)68;Q(12,9)69;Q(12,10)70;Q(12,11)71;Q(12,12)72;Q(12,13)73;Q(12,14)74;Q(12,15)75;Q(13,11)76;Q(13,12)77;Q(13,13)78;Q(13,14)79;Q(13,15)80;Q(14,11)81;Q(14,12)82;Q(14,13)83;Q(14,14)84;Q(14,15)85;Q(14,16)86;Q(14,17)87;Q(15,13)88;Q(15,14)89;Q(15,15)90;Q(15,16)91;Q(15,17)92;Q(16,13)93;Q(16,14)94;Q(16,15)95;Q(16,16)96;Q(16,17)97;Q(16,18)98;Q(16,19)99;Q(17,15)100;Q(17,16)101;Q(17,17)102;Q(17,18)103;Q(17,19)104;Q(18,15)105;Q(18,16)106;Q(18,17)107;Q(18,18)108;Q(18,19)109;Q(18,20)110;Q(18,21)111;Q(19,17)112;Q(19,18)113;Q(19,19)114;Q(19,20)115;Q(19,21)116;Q(20,17)117;Q(20,18)118;Q(20,19)119;Q(20,20)120;Q(21,19)121;Q(21,20)122;R_14_26_5_15_28_42_0_6_17_30_44_1_8_19_32_46_9_21_34_41_48_52_50_54_58_56_60_64_62_66_70_68_72_76_74_78_82_80_84_88_86_90_94_92_96_100_98_102_106_104_108_112_110_114_118_116_120_121_27_29_45_20_47_53_55_65_67_77_79_89_91_101_103_113_115_122;RX_22_23_38_4_13_39_49_57_59_61_69_71_73_81_83_85_93_95_97_105_107_109_117_119;POLYGON(0,0,1,0.25)52_58_54_48;POLYGON(0,0,1,0.25)54_60_56_50;POLYGON(0,0,1,0.25)64_70_66_60;POLYGON(0,0,1,0.25)66_72_68_62;POLYGON(0,0,1,0.25)76_82_78_72;POLYGON(0,0,1,0.25)78_84_80_74;POLYGON(0,0,1,0.25)88_94_90_84;POLYGON(0,0,1,0.25)90_96_92_86;POLYGON(0,0,1,0.25)100_106_102_96;POLYGON(0,0,1,0.25)102_108_104_98;POLYGON(0,0,1,0.25)112_118_114_108;POLYGON(0,0,1,0.25)114_120_116_110;POLYGON(0,0,1,0.25)121_120;POLYGON(1,0,0,0.25)14_5;POLYGON(1,0,0,0.25)26_42;POLYGON(1,0,0,0.25)14_26_28_15;POLYGON(1,0,0,0.25)5_0;POLYGON(1,0,0,0.25)1_9;POLYGON(1,0,0,0.25)5_42_44_6;POLYGON(1,0,0,0.25)0_44_46_1;POLYGON(1,0,0,0.25)8_32_34_9;POLYGON(1,0,0,0.25)34_41;POLYGON(1,0,0,0.25)32_46_52_48;POLYGON(1,0,0,0.25)48_54_50_41;POLYGON(1,0,0,0.25)52_58;POLYGON(1,0,0,0.25)58_64_60_54;POLYGON(1,0,0,0.25)60_66_62_56;POLYGON(1,0,0,0.25)70_76_72_66;POLYGON(1,0,0,0.25)72_78_74_68;POLYGON(1,0,0,0.25)64_70;POLYGON(1,0,0,0.25)76_82;POLYGON(1,0,0,0.25)88_94;POLYGON(1,0,0,0.25)82_88_84_78;POLYGON(1,0,0,0.25)84_90_86_80;POLYGON(1,0,0,0.25)94_100_96_90;POLYGON(1,0,0,0.25)96_102_98_92;POLYGON(1,0,0,0.25)106_112_108_102;POLYGON(1,0,0,0.25)108_114_110_104;POLYGON(1,0,0,0.25)118_121_120_114;POLYGON(1,0,0,0.25)112_118;POLYGON(1,0,0,0.25)100_106;POLYGON(1,0,0,0.25)50_56;POLYGON(1,0,0,0.25)62_68;POLYGON(1,0,0,0.25)74_80;POLYGON(1,0,0,0.25)86_92;POLYGON(1,0,0,0.25)98_104;POLYGON(1,0,0,0.25)110_116;TICK;CX_22_27_23_29_38_45_4_1_13_20_39_47_49_41_57_52_59_54_61_56_69_64_71_66_73_68_81_76_83_78_85_80_93_88_95_90_97_92_105_100_107_102_109_104_117_112_119_114_48_53_50_55_60_65_62_67_72_77_74_79_84_89_86_91_96_101_98_103_108_113_110_115_120_122;R_16_7_31;RX_2_11_3_24;TICK;CX_22_14_27_15_2_5_11_16_23_28_29_17_3_7_24_31_38_30_45_46_4_9_13_8_20_19_39_32_47_48_49_50_57_58_59_60_61_62_69_70_71_72_73_74_81_82_83_84_85_86_93_94_95_96_97_98_105_106_107_108_109_110_117_118_119_120_52_53_54_55_64_65_66_67_76_77_78_79_88_89_90_91_100_101_102_103_112_113_114_115_121_122;R_43_18_33_40;RX_37_12_25_51_63_75_87_99_111;TICK;CX_22_26_27_28_2_0_11_5_16_6_23_15_29_30_37_43_3_1_7_8_12_18_24_17_31_19_38_44_45_32_13_9_20_21_25_33_39_46_47_52_49_48_51_50_59_58_61_60_63_62_71_70_73_72_75_74_83_82_85_84_87_86_95_94_97_96_99_98_107_106_109_108_111_110_119_118_41_40_54_53_56_55_66_65_68_67_78_77_80_79_90_89_92_91_102_101_104_103_114_113_116_115;M_122;MX_4_57_69_81_93_105_117;RX_10;DT(21,20,0)rec[-8];TICK;CX_10_14_22_27_11_15_16_17_23_29_37_28_43_42_3_0_7_6_12_8_18_19_24_30_31_32_38_45_13_20_25_21_33_34_39_47_49_54_51_56_59_64_61_66_63_68_71_76_73_78_75_80_83_88_85_90_87_92_95_100_97_102_99_104_107_112_109_114_111_116_119_121_48_40_58_53_60_55_70_65_72_67_82_77_84_79_94_89_96_91_106_101_108_103_118_113_120_115;MX_2;RX_36_35;TICK;CX_10_5_36_42_11_16_37_30_43_44_3_7_12_6_18_17_24_31_25_19_33_32_35_41;M_27_29_45_20_47_40_53_55_65_67_77_79_89_91_101_103_113_115;MX_22_23_38_13_39_49_51_59_61_63_71_73_75_83_85_87_95_97_99_107_109_111_119;DT(6,1,1)rec[-41];DT(6,3,1)rec[-40];DT(8,5,1)rec[-39];DT(4,7,1)rec[-38];DT(8,7,1)rec[-37];DT(7,8,1)rec[-36];DT(9,8,1)rec[-35];DT(9,10,1)rec[-34];DT(11,10,1)rec[-33];DT(11,12,1)rec[-32];DT(13,12,1)rec[-31];DT(13,14,1)rec[-30];DT(15,14,1)rec[-29];DT(15,16,1)rec[-28];DT(17,16,1)rec[-27];DT(17,18,1)rec[-26];DT(19,18,1)rec[-25];DT(19,20,1)rec[-24];TICK;CX_36_26_37_43_12_18_25_33_35_34;M_16_7_31;MX_10_11_3_24;R_27_29_45_20_47_53_55_65_67_77_79_89_91_101_103_113_115_122;RX_22_23_38_4_13_39_49_57_59_61_69_71_73_81_83_85_93_95_97_105_107_109_117_119;DT(4,3,2)rec[-7];DT(2,5,2)rec[-6];DT(6,5,2)rec[-5];TICK;CX_22_27_23_29_38_45_4_1_13_20_39_47_49_41_57_52_59_54_61_56_69_64_71_66_73_68_81_76_83_78_85_80_93_88_95_90_97_92_105_100_107_102_109_104_117_112_119_114_48_53_50_55_60_65_62_67_72_77_74_79_84_89_86_91_96_101_98_103_108_113_110_115_120_122;M_43_18_33;MX_36_37_12_25_35;R_16_7_31;RX_2_11_3_24;DT(8,3,3)rec[-8];DT(4,5,3)rec[-7];DT(6,7,3)rec[-6];TICK;CX_22_14_27_15_2_5_11_16_23_28_29_17_3_7_24_31_38_30_45_46_4_9_13_8_20_19_39_32_47_48_49_50_57_58_59_60_61_62_69_70_71_72_73_74_81_82_83_84_85_86_93_94_95_96_97_98_105_106_107_108_109_110_117_118_119_120_52_53_54_55_64_65_66_67_76_77_78_79_88_89_90_91_100_101_102_103_112_113_114_115_121_122;R_43_18_33_40;RX_37_12_25_51_63_75_87_99_111;TICK;CX_22_26_27_28_2_0_11_5_16_6_23_15_29_30_37_43_3_1_7_8_12_18_24_17_31_19_38_44_45_32_13_9_20_21_25_33_39_46_47_52_49_48_51_50_59_58_61_60_63_62_71_70_73_72_75_74_83_82_85_84_87_86_95_94_97_96_99_98_107_106_109_108_111_110_119_118_41_40_54_53_56_55_66_65_68_67_78_77_80_79_90_89_92_91_102_101_104_103_114_113_116_115;M_122;MX_4_57_69_81_93_105_117;RX_10;DT(21,20,4)rec[-8]_rec[-73];DT(1,7,4)rec[-7]_rec[-72];DT(10,7,4)rec[-6]_rec[-71];DT(12,9,4)rec[-5]_rec[-70];DT(14,11,4)rec[-4]_rec[-69];DT(16,13,4)rec[-3]_rec[-68];DT(18,15,4)rec[-2]_rec[-67];DT(20,17,4)rec[-1]_rec[-66];TICK;CX_10_14_22_27_11_15_16_17_23_29_37_28_43_42_3_0_7_6_12_8_18_19_24_30_31_32_38_45_13_20_25_21_33_34_39_47_49_54_51_56_59_64_61_66_63_68_71_76_73_78_75_80_83_88_85_90_87_92_95_100_97_102_99_104_107_112_109_114_111_116_119_121_48_40_58_53_60_55_70_65_72_67_82_77_84_79_94_89_96_91_106_101_108_103_118_113_120_115;MX_2;RX_36_35;DT(1,3,5)rec[-1]_rec[-66];TICK;CX_10_5_36_42_11_16_37_30_43_44_3_7_12_6_18_17_24_31_25_19_33_32_35_41;M_27_29_45_20_47_40_53_55_65_67_77_79_89_91_101_103_113_115;MX_22_23_38_13_39_49_51_59_61_63_71_73_75_83_85_87_95_97_99_107_109_111_119;DT(6,1,6)rec[-41];DT(6,3,6)rec[-40];DT(8,5,6)rec[-39];DT(4,7,6)rec[-38];DT(8,7,6)rec[-37];DT(7,8,6)rec[-36]_rec[-101];DT(9,8,6)rec[-35]_rec[-100];DT(9,10,6)rec[-34]_rec[-99];DT(11,10,6)rec[-33]_rec[-98];DT(11,12,6)rec[-32]_rec[-97];DT(13,12,6)rec[-31]_rec[-96];DT(13,14,6)rec[-30]_rec[-95];DT(15,14,6)rec[-29]_rec[-94];DT(15,16,6)rec[-28]_rec[-93];DT(17,16,6)rec[-27]_rec[-92];DT(17,18,6)rec[-26]_rec[-91];DT(19,18,6)rec[-25]_rec[-90];DT(19,20,6)rec[-24]_rec[-89];DT(5,1,6)rec[-23]_rec[-88];DT(5,3,6)rec[-22]_rec[-87];DT(7,5,6)rec[-21]_rec[-86];DT(3,7,6)rec[-20]_rec[-85];DT(7,7,6)rec[-19]_rec[-84];DT(8,9,6)rec[-18]_rec[-83];DT(8,11,6)rec[-17]_rec[-82];DT(10,9,6)rec[-16]_rec[-81];DT(10,11,6)rec[-15]_rec[-80];DT(10,13,6)rec[-14]_rec[-79];DT(12,11,6)rec[-13]_rec[-78];DT(12,13,6)rec[-12]_rec[-77];DT(12,15,6)rec[-11]_rec[-76];DT(14,13,6)rec[-10]_rec[-75];DT(14,15,6)rec[-9]_rec[-74];DT(14,17,6)rec[-8]_rec[-73];DT(16,15,6)rec[-7]_rec[-72];DT(16,17,6)rec[-6]_rec[-71];DT(16,19,6)rec[-5]_rec[-70];DT(18,17,6)rec[-4]_rec[-69];DT(18,19,6)rec[-3]_rec[-68];DT(18,21,6)rec[-2]_rec[-67];DT(20,19,6)rec[-1]_rec[-66];TICK;CX_36_26_37_43_12_18_25_33_35_34;M_16_7_31;MX_10_11_3_24;R_27_29_45_20_47_53_55_65_67_77_79_89_91_101_103_113_115_122;RX_22_23_38_4_13_39_49_57_59_61_69_71_73_81_83_85_93_95_97_105_107_109_117_119;DT(4,3,7)rec[-7];DT(2,5,7)rec[-6];DT(6,5,7)rec[-5];DT(3,1,7)rec[-4]_rec[-69];DT(3,3,7)rec[-3]_rec[-68];DT(1,5,7)rec[-2]_rec[-67];DT(5,5,7)rec[-1]_rec[-66];TICK;CX_22_27_23_29_38_45_4_1_13_20_39_47_49_41_57_52_59_54_61_56_69_64_71_66_73_68_81_76_83_78_85_80_93_88_95_90_97_92_105_100_107_102_109_104_117_112_119_114_48_53_50_55_60_65_62_67_72_77_74_79_84_89_86_91_96_101_98_103_108_113_110_115_120_122;M_43_18_33;MX_36_37_12_25_35;R_16_7_31;RX_2_11_3_24;DT(8,3,8)rec[-8];DT(4,5,8)rec[-7];DT(6,7,8)rec[-6];DT(7,1,8)rec[-5]_rec[-70];DT(7,3,8)rec[-4]_rec[-69];DT(3,5,8)rec[-3]_rec[-68];DT(5,7,8)rec[-2]_rec[-67];DT(6,9,8)rec[-1]_rec[-66];TICK;CX_22_14_27_15_2_5_11_16_23_28_29_17_3_7_24_31_38_30_45_46_4_9_13_8_20_19_39_32_47_48_49_50_57_58_59_60_61_62_69_70_71_72_73_74_81_82_83_84_85_86_93_94_95_96_97_98_105_106_107_108_109_110_117_118_119_120_52_53_54_55_64_65_66_67_76_77_78_79_88_89_90_91_100_101_102_103_112_113_114_115_121_122;R_43_18_33_40;RX_37_12_25_51_63_75_87_99_111;TICK;CX_22_26_27_28_2_0_11_5_16_6_23_15_29_30_37_43_3_1_7_8_12_18_24_17_31_19_38_44_45_32_13_9_20_21_25_33_39_46_47_52_49_48_51_50_59_58_61_60_63_62_71_70_73_72_75_74_83_82_85_84_87_86_95_94_97_96_99_98_107_106_109_108_111_110_119_118_41_40_54_53_56_55_66_65_68_67_78_77_80_79_90_89_92_91_102_101_104_103_114_113_116_115;M_122;MX_4_57_69_81_93_105_117;RX_10;DT(21,20,9)rec[-8]_rec[-73];DT(1,7,9)rec[-7]_rec[-72];DT(10,7,9)rec[-6]_rec[-71];DT(12,9,9)rec[-5]_rec[-70];DT(14,11,9)rec[-4]_rec[-69];DT(16,13,9)rec[-3]_rec[-68];DT(18,15,9)rec[-2]_rec[-67];DT(20,17,9)rec[-1]_rec[-66];TICK;CX_10_14_22_27_11_15_16_17_23_29_37_28_43_42_3_0_7_6_12_8_18_19_24_30_31_32_38_45_13_20_25_21_33_34_39_47_49_54_51_56_59_64_61_66_63_68_71_76_73_78_75_80_83_88_85_90_87_92_95_100_97_102_99_104_107_112_109_114_111_116_119_121_48_40_58_53_60_55_70_65_72_67_82_77_84_79_94_89_96_91_106_101_108_103_118_113_120_115;MX_2;RX_36_35;DT(1,3,10)rec[-1]_rec[-66];TICK;CX_10_5_36_42_11_16_37_30_43_44_3_7_12_6_18_17_24_31_25_19_33_32_35_41;M_27_29_45_20_47_40_53_55_65_67_77_79_89_91_101_103_113_115;MX_22_23_38_13_39_49_51_59_61_63_71_73_75_83_85_87_95_97_99_107_109_111_119;DT(6,1,11)rec[-41];DT(6,3,11)rec[-40];DT(8,5,11)rec[-39];DT(4,7,11)rec[-38];DT(8,7,11)rec[-37];DT(7,8,11)rec[-36]_rec[-101];DT(9,8,11)rec[-35]_rec[-100];DT(9,10,11)rec[-34]_rec[-99];DT(11,10,11)rec[-33]_rec[-98];DT(11,12,11)rec[-32]_rec[-97];DT(13,12,11)rec[-31]_rec[-96];DT(13,14,11)rec[-30]_rec[-95];DT(15,14,11)rec[-29]_rec[-94];DT(15,16,11)rec[-28]_rec[-93];DT(17,16,11)rec[-27]_rec[-92];DT(17,18,11)rec[-26]_rec[-91];DT(19,18,11)rec[-25]_rec[-90];DT(19,20,11)rec[-24]_rec[-89];DT(5,1,11)rec[-23]_rec[-88];DT(5,3,11)rec[-22]_rec[-87];DT(7,5,11)rec[-21]_rec[-86];DT(3,7,11)rec[-20]_rec[-85];DT(7,7,11)rec[-19]_rec[-84];DT(8,9,11)rec[-18]_rec[-83];DT(8,11,11)rec[-17]_rec[-82];DT(10,9,11)rec[-16]_rec[-81];DT(10,11,11)rec[-15]_rec[-80];DT(10,13,11)rec[-14]_rec[-79];DT(12,11,11)rec[-13]_rec[-78];DT(12,13,11)rec[-12]_rec[-77];DT(12,15,11)rec[-11]_rec[-76];DT(14,13,11)rec[-10]_rec[-75];DT(14,15,11)rec[-9]_rec[-74];DT(14,17,11)rec[-8]_rec[-73];DT(16,15,11)rec[-7]_rec[-72];DT(16,17,11)rec[-6]_rec[-71];DT(16,19,11)rec[-5]_rec[-70];DT(18,17,11)rec[-4]_rec[-69];DT(18,19,11)rec[-3]_rec[-68];DT(18,21,11)rec[-2]_rec[-67];DT(20,19,11)rec[-1]_rec[-66];TICK;CX_36_26_37_43_12_18_25_33_35_34;M_16_7_31;MX_10_11_3_24;R_20_47_53_55_65_67_77_79_89_91_101_103_113_115_122;RX_22_13_39_49_57_59_61_69_71_73_81_83_85_93_95_97_105_107_109_117_119;DT(4,3,12)rec[-7];DT(2,5,12)rec[-6];DT(6,5,12)rec[-5];DT(3,1,12)rec[-4]_rec[-69];DT(3,3,12)rec[-3]_rec[-68];DT(1,5,12)rec[-2]_rec[-67];DT(5,5,12)rec[-1]_rec[-66];TICK;CX_22_26_13_20_39_47_49_41_57_52_59_54_61_56_69_64_71_66_73_68_81_76_83_78_85_80_93_88_95_90_97_92_105_100_107_102_109_104_117_112_119_114_48_53_50_55_60_65_62_67_72_77_74_79_84_89_86_91_96_101_98_103_108_113_110_115_120_122;M_43_18_33;MX_36_37_12_25_35;R_27_16_29_7;RX_11_23_3;DT(8,3,13)rec[-8];DT(4,5,13)rec[-7];DT(6,7,13)rec[-6];DT(7,1,13)rec[-5]_rec[-70];DT(7,3,13)rec[-4]_rec[-69];DT(3,5,13)rec[-3]_rec[-68];DT(5,7,13)rec[-2]_rec[-67];DT(6,9,13)rec[-1]_rec[-66];TICK;CX_22_27_11_16_23_29_3_7_39_32_47_48_49_50_57_58_59_60_61_62_69_70_71_72_73_74_81_82_83_84_85_86_93_94_95_96_97_98_105_106_107_108_109_110_117_118_119_120_52_53_54_55_64_65_66_67_76_77_78_79_88_89_90_91_100_101_102_103_112_113_114_115_121_122;R_43_18_45_33_40;RX_37_12_24_38_25_51_63_75_87_99_111;TICK;CX_11_5_16_6_23_30_37_43_3_0_7_1_12_18_24_17_38_45_13_19_20_21_25_33_49_48_51_50_59_58_61_60_63_62_71_70_73_72_75_74_83_82_85_84_87_86_95_94_97_96_99_98_107_106_109_108_111_110_119_118_41_40_54_53_56_55_66_65_68_67_78_77_80_79_90_89_92_91_102_101_104_103_114_113_116_115;M_122;MX_57_69_81_93_105_117;R_31;DT(21,20,14)rec[-7]_rec[-72];DT(10,7,14)rec[-6]_rec[-70];DT(12,9,14)rec[-5]_rec[-69];DT(14,11,14)rec[-4]_rec[-68];DT(16,13,14)rec[-3]_rec[-67];DT(18,15,14)rec[-2]_rec[-66];DT(20,17,14)rec[-1]_rec[-65];TICK;CX_22_14_27_15_11_17_37_28_43_30_12_8_24_31_38_44_45_46_13_9_25_19_33_32_49_54_51_56_59_64_61_66_63_68_71_76_73_78_75_80_83_88_85_90_87_92_95_100_97_102_99_104_107_112_109_114_111_116_119_121_48_40_58_53_60_55_70_65_72_67_82_77_84_79_94_89_96_91_106_101_108_103_118_113_120_115;TICK;CX_22_27_11_16_23_15_29_28_37_42_3_8_12_6_18_17_24_19_31_32_38_30_13_20_25_34_39_52;M_40_53_55_65_67_77_79_89_91_101_103_113_115;MX_49_51_59_61_63_71_73_75_83_85_87_95_97_99_107_109_111_119;RX_10_36_2_4_35;DT(7,8,15)rec[-31]_rec[-89];DT(9,8,15)rec[-30]_rec[-88];DT(9,10,15)rec[-29]_rec[-87];DT(11,10,15)rec[-28]_rec[-86];DT(11,12,15)rec[-27]_rec[-85];DT(13,12,15)rec[-26]_rec[-84];DT(13,14,15)rec[-25]_rec[-83];DT(15,14,15)rec[-24]_rec[-82];DT(15,16,15)rec[-23]_rec[-81];DT(17,16,15)rec[-22]_rec[-80];DT(17,18,15)rec[-21]_rec[-79];DT(19,18,15)rec[-20]_rec[-78];DT(19,20,15)rec[-19]_rec[-77];DT(8,9,15)rec[-18]_rec[-71];DT(8,11,15)rec[-17]_rec[-70];DT(10,9,15)rec[-16]_rec[-69];DT(10,11,15)rec[-15]_rec[-68];DT(10,13,15)rec[-14]_rec[-67];DT(12,11,15)rec[-13]_rec[-66];DT(12,13,15)rec[-12]_rec[-65];DT(12,15,15)rec[-11]_rec[-64];DT(14,13,15)rec[-10]_rec[-63];DT(14,15,15)rec[-9]_rec[-62];DT(14,17,15)rec[-8]_rec[-61];DT(16,15,15)rec[-7]_rec[-60];DT(16,17,15)rec[-6]_rec[-59];DT(16,19,15)rec[-5]_rec[-58];DT(18,17,15)rec[-4]_rec[-57];DT(18,19,15)rec[-3]_rec[-56];DT(18,21,15)rec[-2]_rec[-55];DT(20,19,15)rec[-1]_rec[-54];TICK;CX_10_14_36_26_2_5_23_29_37_43_3_7_12_18_24_31_38_45_4_1_25_33_35_41_39_47;TICK;CX_10_5_22_28_36_42_2_0_11_15_23_17_37_44_3_6_12_19_24_30_38_32_4_9_13_8_25_21_35_34_39_46;M_29_43_7_18_31_45_33_47;DT(6,3,16)rec[-8];DT(8,3,16)rec[-7];DT(2,5,16)rec[-6];DT(4,5,16)rec[-5];DT(6,5,16)rec[-4];DT(8,5,16)rec[-3];DT(6,7,16)rec[-2];DT(8,7,16)rec[-1];TICK;MX_10_22_36_2_11_23_37_3_12_24_38_4_13_25_35_39;DT(3,1,17)rec[-16]_rec[-74];DT(5,1,17)rec[-15]_rec[-100];DT(7,1,17)rec[-14]_rec[-67];DT(1,3,17)rec[-13]_rec[-119];DT(3,3,17)rec[-12]_rec[-73];DT(5,3,17)rec[-11]_rec[-99];DT(7,3,17)rec[-10]_rec[-66];DT(1,5,17)rec[-9]_rec[-72];DT(3,5,17)rec[-8]_rec[-65];DT(5,5,17)rec[-7]_rec[-71];DT(7,5,17)rec[-6]_rec[-98];DT(1,7,17)rec[-5]_rec[-126];DT(3,7,17)rec[-4]_rec[-97];DT(5,7,17)rec[-3]_rec[-64];DT(6,9,17)rec[-2]_rec[-63];DT(7,7,17)rec[-1]_rec[-96];TICK;I_41;R_40_53_55_65_67_77_79_89_91_101_103_113_115_122;RX_49_57_51_59_61_69_63_71_73_81_75_83_85_93_87_95_97_105_99_107_109_117_111_119;SQRT_X_15_44_0_42_30_19_34_8_9_21_46_32_6_28_17;TICK;CX_49_41_57_52_59_54_61_56_69_64_71_66_73_68_81_76_83_78_85_80_93_88_95_90_97_92_105_100_107_102_109_104_117_112_119_114_48_53_50_55_60_65_62_67_72_77_74_79_84_89_86_91_96_101_98_103_108_113_110_115_120_122;M_14_26_5_15_28_42_0_6_17_30_44_1_8_19_32_46_9_21_34;OI(1)rec[-2]_rec[-3]_rec[-6]_rec[-7]_rec[-8]_rec[-9]_rec[-10]_rec[-14]_rec[-15]_rec[-18]_rec[-215]_rec[-246];OI(3)rec[-4]_rec[-5]_rec[-6]_rec[-7]_rec[-9]_rec[-10]_rec[-11]_rec[-12]_rec[-214]_rec[-247];TICK;CX_49_50_57_58_59_60_61_62_69_70_71_72_73_74_81_82_83_84_85_86_93_94_95_96_97_98_105_106_107_108_109_110_117_118_119_120_52_53_54_55_64_65_66_67_76_77_78_79_88_89_90_91_100_101_102_103_112_113_114_115_121_122;TICK;CX_49_48_51_50_59_58_61_60_63_62_71_70_73_72_75_74_83_82_85_84_87_86_95_94_97_96_99_98_107_106_109_108_111_110_119_118_41_40_54_53_56_55_66_65_68_67_78_77_80_79_90_89_92_91_102_101_104_103_114_113_116_115;TICK;CX_49_54_51_56_59_64_61_66_63_68_71_76_73_78_75_80_83_88_85_90_87_92_95_100_97_102_99_104_107_112_109_114_111_116_119_121_48_40_58_53_60_55_70_65_72_67_82_77_84_79_94_89_96_91_106_101_108_103_118_113_120_115;TICK;M_40_53_55_65_67_77_79_89_91_101_103_113_115_122;MX_49_57_51_59_61_69_63_71_73_81_75_83_85_93_87_95_97_105_99_107_109_117_111_119;MY_41_48_52_50_54_58_56_60_64_62_66_70_68_72_76_74_78_82_80_84_88_86_90_94_92_96_100_98_102_106_104_108_112_110_114_118_116_120_121;DT(9,8,18)rec[-76]_rec[-150];DT(9,10,18)rec[-75]_rec[-149];DT(11,10,18)rec[-74]_rec[-148];DT(11,12,18)rec[-73]_rec[-147];DT(13,12,18)rec[-72]_rec[-146];DT(13,14,18)rec[-71]_rec[-145];DT(15,14,18)rec[-70]_rec[-144];DT(15,16,18)rec[-69]_rec[-143];DT(17,16,18)rec[-68]_rec[-142];DT(17,18,18)rec[-67]_rec[-141];DT(19,18,18)rec[-66]_rec[-140];DT(19,20,18)rec[-65]_rec[-139];DT(21,20,18)rec[-64]_rec[-158];DT(8,9,18)rec[-63]_rec[-138];DT(10,7,18)rec[-62]_rec[-157];DT(8,11,18)rec[-61]_rec[-137];DT(10,9,18)rec[-60]_rec[-136];DT(10,11,18)rec[-59]_rec[-135];DT(12,9,18)rec[-58]_rec[-156];DT(10,13,18)rec[-57]_rec[-134];DT(12,11,18)rec[-56]_rec[-133];DT(12,13,18)rec[-55]_rec[-132];DT(14,11,18)rec[-54]_rec[-155];DT(12,15,18)rec[-53]_rec[-131];DT(14,13,18)rec[-52]_rec[-130];DT(14,15,18)rec[-51]_rec[-129];DT(16,13,18)rec[-50]_rec[-154];DT(14,17,18)rec[-49]_rec[-128];DT(16,15,18)rec[-48]_rec[-127];DT(16,17,18)rec[-47]_rec[-126];DT(18,15,18)rec[-46]_rec[-153];DT(16,19,18)rec[-45]_rec[-125];DT(18,17,18)rec[-44]_rec[-124];DT(18,19,18)rec[-43]_rec[-123];DT(20,17,18)rec[-42]_rec[-152];DT(18,21,18)rec[-41]_rec[-122];DT(20,19,18)rec[-40]_rec[-121];OI(0)rec[-77]_rec[-78]_rec[-80]_rec[-81]_rec[-83]_rec[-85]_rec[-86]_rec[-88]_rec[-90]_rec[-92]_rec[-94]_rec[-96]_rec[-290]_rec[-291]_rec[-299]_rec[-323]_rec[-324]_rec[-325]_rec[-345];OI(2)rec[-77]_rec[-78]_rec[-79]_rec[-82]_rec[-83]_rec[-87]_rec[-88]_rec[-92]_rec[-93]_rec[-290]_rec[-325];OI(4)rec[-1]_rec[-2]_rec[-3]_rec[-4]_rec[-5]_rec[-6]_rec[-7]_rec[-8]_rec[-9]_rec[-10]_rec[-11]_rec[-12]_rec[-13]_rec[-14]_rec[-15]_rec[-16]_rec[-17]_rec[-18]_rec[-19]_rec[-20]_rec[-21]_rec[-22]_rec[-23]_rec[-24]_rec[-25]_rec[-26]_rec[-27]_rec[-28]_rec[-29]_rec[-30]_rec[-31]_rec[-32]_rec[-33]_rec[-34]_rec[-35]_rec[-36]_rec[-37]_rec[-38]_rec[-39]_rec[-78]_rec[-79]_rec[-80]_rec[-81]_rec[-82]_rec[-83]_rec[-84]_rec[-85]_rec[-289]_rec[-291]_rec[-292]_rec[-293]_rec[-299]_rec[-300]_rec[-304]_rec[-305]_rec[-307]_rec[-308]_rec[-310]_rec[-311]_rec[-313]_rec[-314]_rec[-316]_rec[-317]_rec[-319]_rec[-320]_rec[-322]_rec[-323]_rec[-324]_rec[-326]}{The circuit can be found in Crumble}.}
    \label{fig:surface code}
\end{figure*}

In Section~\ref{sec:repetition code}, we examined unfolded distillation in the absence of bit-flip errors. In the proposed scheme, the output magic state is encoded within a repetition code, requiring that no bit-flip errors occur on any of the qubits within the repetition code until the magic state is consumed. In this section, we demonstrate that unfolded distillation is highly robust against bit-flip errors and that, when the output magic state is encoded in a thin surface code, a magic state of the same fidelity as that in Section~\ref{sec:repetition code} can be prepared, even when the noise bias is as low as $\eta \approx 80$, using the same unfolded code. The main idea is that the Hadamard Reed-Muller code can be used not only to detect faulty $X^{-1/4}$ rotations, but also errors on the biased-noise qubits during the rounds of X-type stabilizer measurement. A similar idea was used in Ref.~\cite{litinski2019magic} to reduce the distance of the inner surface codes employed for distillation. Here, the protection granted by the Hadamard Reed-Muller code can be used to detect bit-flip errors on the biased-noise qubits within the unfolded code.

Indeed, in the circuit presented in Figure~\ref{fig:unfolding}, a bit-flip error on a data qubit commutes with the X-type stabilizers and the $X^{-1/4}$ gates, and flips the outcome of the final $Z$ measurement. In the absence of other errors, the bit-flip error is detected by the reconstructed $Z$-type stabilizers of the Hadamard Reed-Muller code and the distillation process is restarted. If there are no phase-flip errors, a minimum of three bit-flip errors on data qubits is required in the unfolded code to induce an undetected logical error. Thus, unfolded distillation can be employed with qubits exhibiting a modest noise bias by merging the unfolded code with a thin surface code using lattice surgery, as illustrated in Figure~\ref{fig:surface code}(a). It can be noted that, during the merging, the weight-2 $ZZ$ surface code stabilizer adjacent to the unfolded code should initially not be measured. Indeed, this stabilizer is merged with $Z$ stabilizers of the Hadamard Reed-Muller code. In practice, this means that to reconstruct the latter and accept or reject the magic state, the Hadamard Reed-Muller stabilizers that are merged need to be multiplied by the $ZZ$ surface code stabilizer when it is measured again.

\begin{figure*}[ht]
    \centering
    \includegraphics[width=\textwidth]{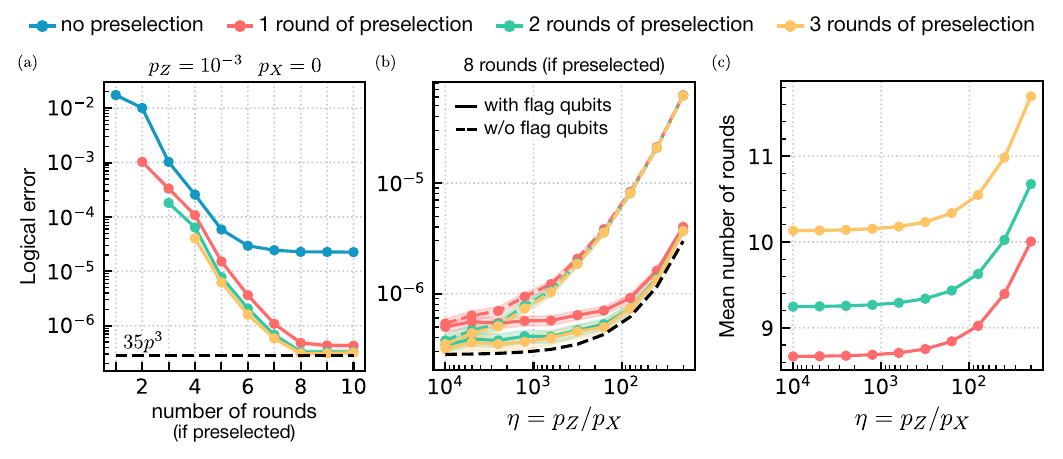}
    \caption{Logical error rate of unfolded distillation using a thin surface code under moderate noise bias. The surface code has a distance $d_X = 3$ and $d_Z = 13$. (a) The logical error is first evaluated considering only phase-flip errors. This enables us to determine the number of $X$-stabilizer measurement rounds and preselection rounds required to achieve a distillation error of $35p^3$, where $p$ is the physical error rate of the $X^{-1/4}$ gate. For a phase-flip error rate of $p_Z = 10^{-3}$, eight rounds of syndrome extraction, including one to two preselection rounds, are sufficient. (b) Keeping the number of rounds fixed at eight, the bit-flip error rate is then varied. The use of flag qubits mitigates the impact of bit-flip errors, and even at a noise bias as low as $\eta \approx 80$, the increase in logical error rate for the magic state remains modest. The analytical dotted curve corresponds to the $35p^3$ limit of distillation, which incorporates both $X^{-1/4}$ gate errors and bit-flip errors during the eight rounds of $X$-type stabilizer measurements, and it shows excellent agreement with the numerical simulations. (c) The mean number of rounds, with a minimum of eight rounds, includes contributions from preselection, postselection due to distillation, the use of flag qubits, and the complementary gap procedure. As noise bias decreases, increased postselection on the flag qubits results in a higher average number of rounds required. In the simulations, $p_Y = 0$ was assumed due to simulation runtime constraints. However, we do not expect the results to change with the inclusion of $Y$ errors.}
    \label{fig:numerics surface code}
\end{figure*}

While the scheme is robust to data qubit bit-flip errors, it is nevertheless sensitive to bit-flip hook errors on the ancilla qubits, which propagate to two bit-flip errors on the data qubits. This effectively reduces the distance of the Hadamard Reed-Muller code against bit-flip errors from 3 to 2. A solution to preserve the distance is the use of flag qubits~\cite{chao2018quantum}, as illustrated in Figure~\ref{fig:surface code}(b). Unlike conventional flag qubit schemes, our approach does not allow for correlated decoding between the flag qubit measurement and subsequent $Z$ stabilizers, as only $X$ stabilizers are present. Thus, it is necessary to post-select on the flag qubits yielding a trivial outcome, and rejecting the magic state preparation otherwise. In the considered noise-bias regime, this postselection increases the time cost only marginally. While merging the unfolded code with a $d_X= 3$ surface code is convenient for the 2D layout, this choice limits the fidelity of the output magic state in the low noise-bias regime. Our proposed solution is to apply complementary gap postselection~\cite{gidney2024yoked} on the $d_X=3$ surface code until the unfolded code is measured and the surface code is grown to $d_X=5$. The complementary gap quantifies the difference in probability between the correction proposed by the decoder and the most likely correction corresponding to a different logical outcome. In the case of decoding via minimum weight perfect matching, this gap can be efficiently computed and is an estimate of the likelihood of the decoder’s proposed correction. This approach ensures that the $d_X = 3$ surface code does not limit the logical fidelity of the magic state. Once again, in the considered noise-bias regime, this complementary gap postselection increases the time cost only marginally. Alternatively, we note that the unfolded code could be merged to a $d_X=5$ thin surface code by slightly modifying the stabilizer geometry. The phase-flip distance of the surface code must be chosen as well such that it does not limit the overall fidelity of the magic state. Here, we found numerically that $d_Z = 13$ is sufficient to meet this requirement.

As in the case of unfolded distillation using a repetition code, a minimum number of stabilizer measurement rounds is required, as stabilizer measurements are noisy. To first evaluate the number of rounds needed, Figure~\ref{fig:numerics surface code}(a) presents the logical error rate of the prepared magic state as a function of the number of $X$ stabilizer measurement rounds, evaluated at a physical error rate of $p_Z = 10^{-3}$ and $p_X = p_Y = 0$. In this case, a minimum of 8 stabilizer rounds (including at least one preselected round) is required to saturate the intrinsic $35p^3$ lower bound set by distillation, more than in the repetition code case, because the longer stabilizer measurements are more susceptible to noise. Preselection, which requires all syndrome measurements to be trivial for a certain number of rounds prior to the transversal $X^{1/4}$ gate, is also necessary, but it can be avoided at the cost of increasing the number of measurement rounds, as in the repetition code case, as detailed in Appendix~\ref{app:Clifford ladder}. Figure~\ref{fig:numerics surface code}(b) shows the logical error as a function of the noise bias $\eta$ after 8 rounds of stabilizer measurements. As expected, in the absence of flag qubits, the logical error rate increases significantly once the noise bias is below $\eta = 10^4$, whereas with flag qubits, it remains relatively stable up to $\eta = 10^2$. Figure~\ref{fig:numerics surface code}(c) shows the average total number of rounds required, including distillation acceptance, preselection, complementary gap postselection and flag postselection. Overall, a magic state with error probability $7 \times 10^{-7}$ can be prepared with fewer than 10 rounds of stabilizer measurements under a noise bias $\eta = 80$. 

\section{Universal set of gate construction}
\label{sec:universal}

In this section, we present a construction of a universal gate set for biased-noise qubits using unfolded distillation. 
The gate set $\{\mathcal{P}_{\ket{0}}, \mathcal{P}_{\ket{+}}, X, Z, \mathrm{CX} \}$ can be implemented either transversally or through lattice surgery on the repetition code~\cite{guillaud2019repetition, gouzien2023performance}, on the phase-flip LDPC codes~\cite{ruiz2025ldpc}, or on the thin surface code~\cite{chamberland2022building}. As presented in Sections~\ref{sec:repetition code} and~\ref{sec:surface_code}, the magic state $\ket{X^{1/4}}$ can be prepared with a logical error rate of approximately $10^{-7}$, below which unfolded distillation must be concatenated with a conventional magic state distillation protocol, such as the magic state factories described in Refs.~\cite{litinski2019magic,gidney2019efficient}. To complete the universal gate set, the Hadamard gate is required, along with the $S$ or $X^{1/2}$ gate, which is necessary to perform the Clifford correction following the teleportation of the magic state, as illustrated in Figure~\ref{fig:T_teleport}. While the Hadamard and S gates can be implemented efficiently on surface codes via code deformation~\cite{brown2017poking,fowler2012surface}, this approach does not generalize to the repetition code, and it remains unclear how they extend to the thin surface code. As a result, a $\ket{S} = \ket{+i}$ state is required to implement the $S$ or $X^{1/2}$ gate. Similarly, the Hadamard gate requires a gate teleportation protocol for its implementation. One approach is to teleport it using a $\ket{+}$ state, provided a $\ket{CZ} = \tfrac{1}{2}(\ket{00} + \ket{01} + \ket{10} - \ket{11})$ resource state is available (see Figure 1 of Ref.~\cite{mantri2017universality}), or alternatively, it can be implemented using three copies of a $\ket{S}$ state (see Appendix J.2 of Ref.~\cite{chamberland2022building}).

\begin{figure}[t]
    \centering
    \includegraphics[width=0.48\textwidth]{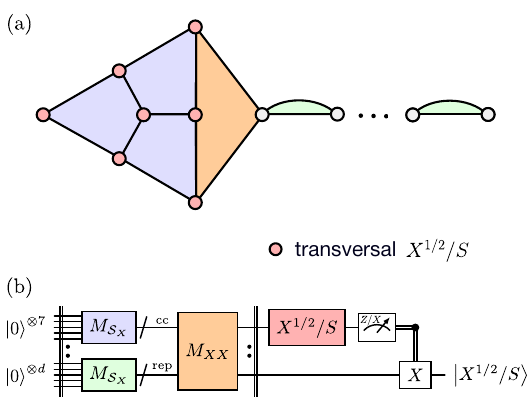}
    \caption{Preparation of the $|S/X^{1/2}\rangle$ state via unfolded distillation. This protocol adapts unfolded distillation by merging a repetition code with a 2D color code~\cite{bombin2006topological} (here a Steane code~\cite{steane1996error}) instead of a quantum 3D Reed-Muller code, as shown in (a). As in the non-Clifford distillation protocol, all qubits are initialized in the $\ket{0}$ state, followed by several rounds of $X$-type stabilizer measurements. Since all Clifford gates are transversal on 2D color codes, either a transversal $X^{1/2}$ or $S$ gate can be applied, followed by measurement in the $Z$ or $X$ basis, respectively. After postselection based on the stabilizers of the 2D color code, the repetition code is projected into the state $\ket{S} = \ket{+i}$ or $|X^{1/2}\rangle = \ket{-i}$. We note that higher-fidelity states can be obtained by using a higher-distance 2D color code, while remaining within a 2D layout.}
    \label{fig:S gate}
\end{figure}

\begin{figure}[t]
    \centering
    \includegraphics[width=0.48\textwidth]{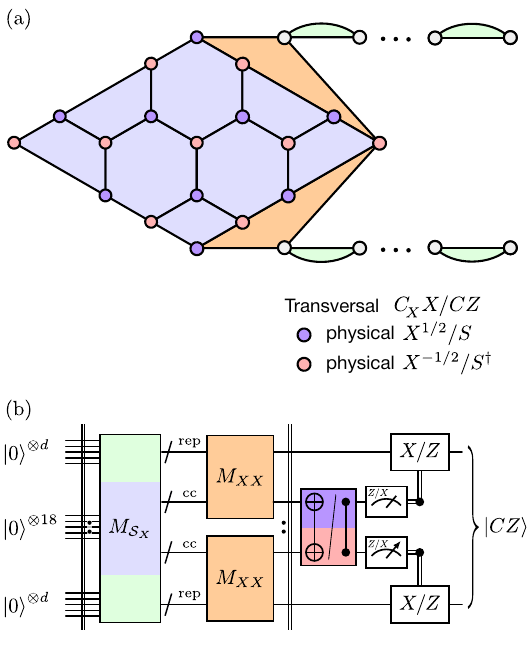}
    \caption{Preparation of the $|CZ\rangle$ state via unfolded distillation. This protocol adapts unfolded distillation by merging two repetition codes with a 2D color code encoding two logical qubits~\cite{kubica2015unfolding}, that supports a transversal $C_XX /CZ$ gate, as shown in (a). $C_XX$ denotes the operation in which the target qubit undergoes an $X$ operation conditioned on the control qubit being in the $+1$ eigenstate of the $X$ basis. As in the non-Clifford distillation protocol, all qubits are initialized in the $\ket{0}$ state, followed by several rounds of $X$-type stabilizer measurements. Since all Clifford gates are transversal on 2D color codes, either a transversal $C_XX$ or $CZ$ gate can be applied, followed by measurement in the $Z$ or $X$ basis, respectively. Thanks to the structure of the 2D color code, the logical $C_XX$ or $CZ$ are implemented by applying physical $X^{\pm 1/2}$ or $S/S^\dagger$ gates, respectively. After postselection based on the stabilizers of the 2D color code, the repetition codes are projected into the state $|CZ\rangle$. We note that higher-fidelity states can be obtained by using a higher-distance 2D color code, while remaining within a 2D layout.}
    \label{fig:CZ gate}
\end{figure}

\begin{figure}[t]
    \centering
    \includegraphics[width=0.48\textwidth]{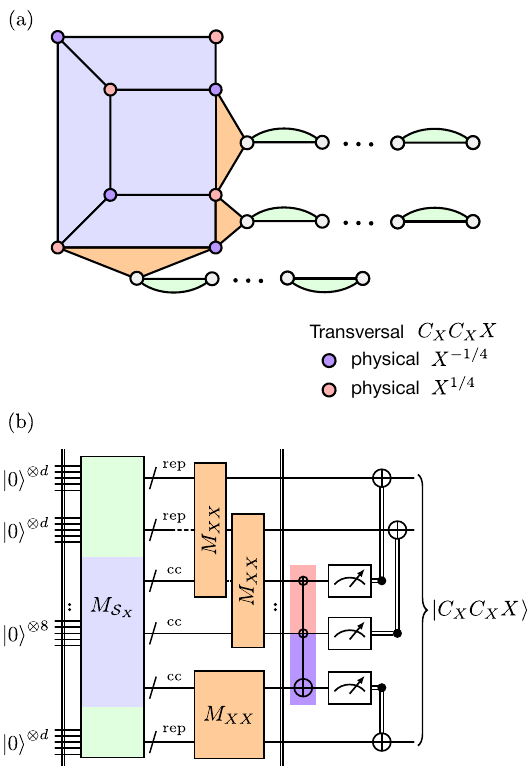}
    \caption{Preparation of the $|C_XC_X X\rangle = C_XC_X X \ket{000}$ magic state via unfolded distillation. This protocol adapts unfolded distillation by merging three repetition codes with the 3D Hadamard $[[8,3,2]]$ color code, which encodes three logical qubits~\cite{kubica2015unfolding} and supports a transversal $C_XC_XX$ gate, as shown in (a). $C_XC_XX$ denotes the operation in which the target qubit undergoes an $X$ operation conditioned on the two control qubits being in the $+1$ eigenstate of the $X$ basis. As in the unfolded distillation protocol for the $|X^{1/4}\rangle$ state, all qubits are initialized in the $\ket{0}$ state, followed by several rounds of $X$-type stabilizer measurements. Then, a transversal $C_XC_XX$ gate is applied, followed by a measurement in the $Z$ basis. After postselection based on the stabilizers of the 3D color code, the repetition codes are projected into the magic state $|C_XC_X X\rangle$.}
    \label{fig:CCZ gate}
\end{figure}

Unfolded distillation provides a cost-effective approach to overcome these limitations, as it can be adapted to prepare the state $\ket{X^{1/2}} = X^{1/2} \ket{0} = \ket{-i}$ or $\ket{S} = \ket{+i}$, thereby enabling the implementation of the $X^{1/2}$ or $S$ gate. It can also be tailored to produce a $\ket{CZ}$ state, thus providing efficient implementation of both the $CZ$ and Hadamard gates using a single magic state. Indeed, instead of employing a Reed-Muller code, a 2D color code~\cite{bombin2006topological} that possesses transversal Clifford gates can be used. Figure~\ref{fig:S gate}(a) shows the preparation of a $\ket{X^{1/2}}$ or $\ket{S}$ state with unfolded distillation using a 2D color code featuring transversal $X^{1/2}$ and $S$ gates. The protocol proceeds by initializing the data qubits of the 2D color code in the $\ket{0}$ state and measuring the $X$ stabilizers for multiple rounds. In contrast to unfolded distillation for a non-Clifford gate, here either a transversal $X^{1/2}$ or $S$ gate can be applied, followed by a measurement of the data qubits in the $Z$ basis or $X$ basis, respectively, to reconstruct the color code stabilizers, as shown in Figure~\ref{fig:S gate}(b). If the color code is merged with a repetition code, the latter will be respectively in the state $\ket{X^{1/2}}$ or $\ket{S}$ at the end of the protocol. Alternatively, the color code could be merged with a thin surface code to output the magic state in the latter, equivalently to what was presented in Section~\ref{sec:surface_code}. We note that while a distance-$3$ 2D color code would yield a state with an error probability of $7p^3$, where $p$ is the error probability of the $X^{1/2}/S$ gate, the 2D color code can easily be extended to prepare states with higher fidelity, for instance a distance-5 2D color code would yield a state with error probability~$36p^5$.

This approach can also be extended to prepare $\ket{CZ}$ states by employing a 2D color code that encodes two logical qubits and supports a transversal $CZ$ or $C_X X$ gate (where the target qubit undergoes an $X$ operation conditioned on the control qubit being in the $+1$ eigenstate of the $X$ basis), as illustrated in Figure~\ref{fig:CZ gate}(a). Interestingly, the structure of the color code allows for the implementation of a logical $CZ$ or $C_X X$ gate, as shown in Figure~\ref{fig:CZ gate}(b), using only physical $X^{\pm 1/2}$ or $S/S^\dagger$ gates, respectively. This, in turn, enables efficient realization of both the $CZ$ and Hadamard gates. Finally, unfolded distillation can be used to prepare the magic state $\ket{C_XC_XX} = C_XC_XX \ket{000}$, where $C_XC_XX$ denotes an $X$ operation on the third qubit conditioned on qubits 1 and 2 being in the $+1$ eigenstate of the $X$ basis, \textit{i.e} $C_XC_X X = (H \otimes H \otimes I) CCX (H \otimes H \otimes I)$. This non-Clifford state enables the direct implementation of a Toffoli gate, which would otherwise require four $T$ gates~\cite{selinger2013quantum}. Figure~\ref{fig:CCZ gate} illustrates how the Hadamard $[[8,3,2]]$ 3D color code~\cite{kubica2015unfolding}, which supports a transversal $C_XC_X X$ operation, can be unfolded and employed within the unfolded distillation protocol to prepare the state $\ket{C_XC_X X}$. Similarly to the protocol used for preparing the $\ket{CZ}$ state, this method requires only physical $X^{\pm 1/4}$ gates. This scheme only produces a magic state with an error probability of $28p^2$, where $p$ denotes the physical error rate of the $X^{\pm 1/4}$ gates, and it remains an open question whether the construction can be generalized to enable higher-fidelity preparation of the $\ket{C_XC_X X}$ state.

\section{Experimental Proposal with a hybrid architecture}
\label{sec:hybrid}

In this section, we discuss how unfolded distillation can be implemented in the case of biased-noise dissipative cat qubits. Cat qubits encode information in coherent states of a harmonic oscillator~\cite{mirrahimi2014dynamically}. Two-photon dissipation enables the stabilization of the harmonic oscillator state within the cat qubit codespace. The bit-flip error, a transition from one coherent state to the other, is then exponentially suppressed with the average number of photons $\bar n$ in the coherent states. The phase-flip rate increases linearly with $\bar{n}$, allowing cat qubits to achieve very low bit-flip error rates without significantly compromising their phase-flip performance. Cat qubits can implement the CNOT, $\mathcal{P}_{\ket{0}}$, and $M_Z$ operations in a bias-preserving manner~\cite{guillaud2019repetition}, all of which are essential for executing unfolded magic state distillation.

\begin{figure}[t]
    \centering
    \includegraphics[width=0.5\textwidth]{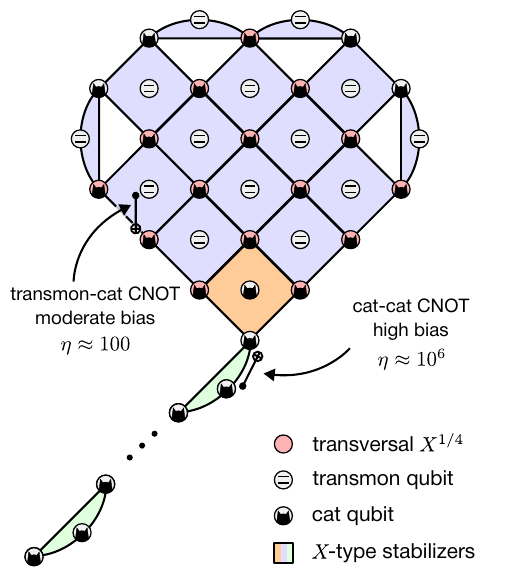}
    \caption{Unfolded distillation with a hybrid cat-transmon architecture. Because the Hadamard Reed-Muller code used to distill $| X^{1/4} \rangle$ magic state also provides protection against bit-flip errors on data cat qubits within the unfolded code, adopting a hybrid cat-transmon architecture proves advantageous. Replacing the ancilla cat qubits in the unfolded code with transmon qubits enables higher-fidelity $X^{-1/4}$ gates, essential for distillation, while the reduced noise bias of the cat-transmon CNOT gate is mitigated by the bit-flip protection offered by the Hadamard Reed-Muller code. This hybrid approach does not prevent the use of a fully cat-based architecture in the output repetition code, which hosts the final magic state, to minimize the qubit overhead.}
    \label{fig:hybrid}
\end{figure}

In the remainder of this section, we discuss the implementation of the $X^{-1/4}$ gate, the final component required to realize unfolded distillation. The physical error rate is a critical input to the distillation protocol, as the output error scales with the cube of the input error. It is therefore crucial that the $X^{-1/4}$ operation is performed with high fidelity. One possibility is to perform a holonomic gate~\cite{albert2016holonomic}. This involves first mapping the cat qubit manifold onto the $\ket{0/1}$ Fock state manifold, and then performing the desired rotation within this manifold. However, with the currently envisioned implementation, such protocols are limited to an error probability of a few percent. An alternative approach could involve using a hybrid transmon-cat qubit architecture for implementing the unfolded code, while preserving a full cat qubit repetition code architecture for the output magic state. Indeed, since unfolded distillation can tolerate a high level of bit-flip errors, an architecture that combines cat qubits as data qubits and transmon qubits as ancilla qubits could be advantageous~\cite{hann2024hybrid, putterman2025hardware}. Such a setup trades a stronger noise bias for increased gate fidelity, benefiting from the simpler interaction between cat and transmon qubits compared to cat-cat interactions. The $X^{-1/4}$ gate on the cat qubit could then be implemented through the ancilla transmon qubit using a SNAP gate by applying a $e^{i \pi/4}$ phase to the odd Fock states~\cite{heeres2015cavity, krastanov2015universal} or through quantum optimal control. Such protocols have been demonstrated to achieve percent to sub-percent infidelities (see, for instance, Refs.~\cite{ heeres2017implementing, eickbusch2022fast, mehta2025bias}). Figure~\ref{fig:hybrid} illustrates a hybrid architecture in which cat qubits are used as data qubits and transmon qubits serve as ancilla qubits within the unfolded code (flag qubits could also be added, as described in Section~\ref{sec:surface_code}), while ancilla qubits in the output repetition code are cat qubits. This enables leveraging the high-fidelity gates of a harmonic oscillator coupled to a transmon qubit without compromising the logical bit-flip error rate, thanks to the bit-flip protection offered by the Hadamard Reed-Muller code. At the same time, a repetition code is used for the output magic state to minimize the overhead.

\section{Conclusion}

We have introduced a highly economical scheme for non-Clifford magic state preparation that leverages biased-noise qubits. The unfolded distillation scheme can prepare a $\ket{X^{1/4}}$ magic state with a logical error rate of $3 \times 10^{-7}$, using only 53 qubits and 5.5 error correction rounds, under a noise bias of $\eta \gtrsim 5 \times 10^6$ and a phase-flip noise rate of 0.1\%. Our approach surpasses existing constructions for biased-noise qubits, especially at more moderate phase-flip error rates such as $p_Z = 5 \times 10^{-3}$, while requiring only a two-dimensional layout and two-qubit gates. Moreover, the method remains effective even at substantially lower noise bias of $\eta \gtrsim 80$, where a magic state with a logical error rate of $7 \times 10^{-7}$ can be prepared using 175 biased-noise qubits and 9.6 rounds of error correction. Relative to state-of-the-art schemes designed for unbiased qubits, our approach lowers the circuit volume required by more than an order of magnitude. This overhead reduction stems from the fact that biased-noise qubits, which predominantly exhibit a single type of error, allow the spatial dimensionality of magic state distillation to be reduced from 3D to 2D. This then enables distillation to be performed directly at the physical level. Extending the code distance of unfolded construction for non-Clifford gates to reach even higher target fidelities is a promising direction for future work.

\section*{Code availability}
The code used for numerical simulations is available at~\cite{CodeGithub}.

\section*{Acknowledgement}

The authors would like to thank Élie Gouzien, Louis Paletta, Vivien Londe and Pierre Guilmin for insightful discussions regarding the proposal, and Ronan Gautier for valuable input on the implementation with cat qubits. This work was supported by the Plan France 2030 under project ANR-22-PETQ-0006. 

\appendix 

\section{Architecture comparison}
\label{app:architecture comparison}

\textit{Bottom up}. We focus on the scheme presented in Ref.~\cite{gouzien2023performance} Appendix E.4, building on the \textit{Bottom-up} scheme presented in Ref.~\cite{chamberland2022building}. The simulations can be found at~\cite{CodeGithub}. The scheme consists of measuring the logical operator $X_1(CX)_{23}$ on 3 repetition codes using a GHZ state (see Section~\ref{sec:state of the art}B) and thus requires $3(2d-1)+d$ qubits. The values presented in Table~\ref{tab:overhead} correspond to the case $d=7$. The number of cycles is set to $5$ for $p_Z = 10^{-3}$ and $6$ for $p_Z = 5 \times 10^{-3}$, while the acceptance rate is $40\%$ and $0.4\%$, respectively. We now compute the required noise bias. A bit-flip error at any location on the three repetition codes corresponds to a logical bit-flip. Therefore, the logical bit-flip rate can be approximated by
\begin{align}
p_{X_L} &= n_{\text{rounds}} \Big[ 
    n_{\text{data}} \left(2p^\text{idle}_X + 2p^\text{CNOT}_{X_t}\right) \notag \\
&+ n_{\text{ancilla}} \left(p^\text{CNOT}_{X_c} + p^\text{CNOT}_{X_cX_t}\right) 
\Big] \notag \\
&= n_{\text{rounds}} \Big[ 
    d \left(2p_X + \frac{2}{3}p_X\right) 
    + (d-1) \left(\frac{p_X}{3} + \frac{p_X}{3}\right) 
\Big]
\label{eq:logical bitflip}
\end{align}
where $p^\text{CNOT}_{X_t}$, $p^\text{CNOT}_{X_c}$ denote the probabilities of a bit-flip error on the target and control qubit of the CNOT gate, respectively, and $p^\text{CNOT}_{X_cX_t}$ denotes the probability of a correlated bit-flip error. We then consider a uniform circuit-level bit-flip noise of strength $p_X$. The noise bias is defined as $\eta = p_Z / p_X$, and we choose $p_X$ such that the logical bit-flip rate $p_{X_L}$ is approximately one order of magnitude smaller than the logical phase-flip rate $p_{Z_L}$. Specifically, for $p_Z = 10^{-3}$, with $n_{\text{rounds}} = 5$ and $d = 7$, setting $p_X = 2 \times 10^{-11}$ yields a logical bit-flip error rate of $p_{X_L} = 2.3 \times 10^{-9}$ for each of the three repetition codes. For $p_Z = 5 \times 10^{-3}$, with $n_{\text{rounds}} = 6$ and $d = 7$, setting $p_X = 10^{-8}$ yields a logical bit-flip error rate of $p_{X_L} = 1.4 \times 10^{-6}$ for each of the three repetition codes.

\textit{Magic state cultivation}. Figure~1 of Ref.~\cite{gidney2024magic} reports the volume cost, while the qubit count is provided in Ref.~\cite{gidney2024magicZenodo}. We estimate the number of rounds by dividing the volume by the number of qubits.

\textit{Unfolded distillation}. The logical error rates, as well as the number of qubits and rounds, are given in Sections~\ref{sec:repetition code} and \ref{sec:surface_code}, and Appendices~\ref{app:Clifford ladder} and \ref{app:high error}. We proceed to compute the noise bias necessary for implementing the repetition code scheme. Since the unfolded code is resilient to bit-flip errors, the dominant contribution to the logical bit-flip error arises from bit-flip errors occurring within the repetition code. Using equation (\ref{eq:logical bitflip}), for the first unfolded code scheme presented in Table~\ref{tab:overhead}, setting $p_X = 2 \times 10^{-10}$ is sufficient to ensure that the magic state fidelity is not limited by bit-flip errors. Specifically, with  $n_{\text{rounds}} = 5.5$ and $d = 9$, the resulting logical bit-flip error rate is $p_{X_L} = 3.2 \times 10^{-8}$, which is an order of magnitude below the magic state logical error. For the second scheme, setting $n_{\text{rounds}} = 12$, $d = 9$ and $p_X = 2\times 10^{-11}$ yields $p_{X_L} = 7.0 \times 10^{-9}$. For the third scheme, setting $n_{\text{rounds}} = 17.7$, $d = 11$ and $p_X = 4 \times 10^{-9}$ yields $p_{X_L} = 2.5 \times 10^{-6}$. For the fourth scheme, setting $n_{\text{rounds}} = 15$, $d = 11$ and $p_X = 2 \times 10^{-9}$ yields $p_{X_L} = 1.1 \times 10^{-6}$. The noise bias is then calculated as $\eta = p_Z / p_X$.

\section{CNOT scheduling}
\label{app:Cnot ordering}

\begin{figure}[t]
    \centering
    \includegraphics[width=0.5\textwidth]{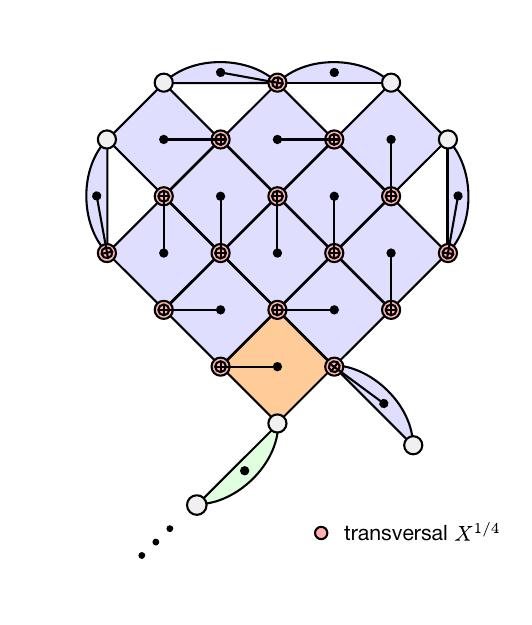}
    \caption{CNOT scheduling for the unfolding distillation protocol. In this protocol, an $X^{1/4}$ gate is applied transversally to all qubits shown in red. However, any phase-flip error occurring before this gate can be transformed into a bit-flip error afterward, potentially leading to a logical error in the distillation process. To minimize this possibility, it is crucial to reduce the time interval between the application of the $X^{-1/4}$ gate to a red qubit and the final CNOT involved in measuring the $X$-type stabilizer on that qubit. To achieve this, it is convenient to introduce an additional data qubit and an ancilla qubit, allowing all CNOTs on the qubits involved in the transversal $X^{1/4}$ gate to be performed simultaneously in the last step of the stabilizer measurements. This synchronization would not be possible without the addition of extra qubits.}
    \label{fig: extra qubit}
\end{figure}

In the scheme described in Section~\ref{sec:repetition code}, it is necessary to determine a CNOT gate schedule such that the gates do not overlap and all stabilizers are measured within four timesteps. Additionally, it is advantageous for every qubit to which the $X^{-1/4}$ gate is applied to be involved in a CNOT interaction immediately beforehand. If this is not the case, the $Z$ idling errors occurring between the CNOT and the $X^{-1/4}$ gate directly contribute to the physical input error rate of the distillation protocol. However, the layout shown in Figure~\ref{fig:unfolding} does not permit all required CNOTs between the ancilla and data qubits supporting the transversal $X^{1/4}$ gate to be performed in a single timestep. While in the scheme from Appendix~\ref{app:Clifford ladder}, the issue can be avoided by simply staggering the $X^{-1/4}$ operations across multiple timesteps, the scheme in Section~\ref{sec:repetition code} requires a different strategy, as $X^{-1/4}$ gates are only applied when the preselection condition is satisfied. One viable solution is to introduce an additional qubit and stabilizer, as illustrated in Figure~\ref{fig: extra qubit}. Thus, by introducing only two additional qubits, it becomes possible to apply a CNOT gate to each qubit involved in the $X^{1/4}$ transversal gate immediately beforehand.

\section{Unfolded distillation without just-in-time decoding or preselection}
\label{app:Clifford ladder}

The unfolded distillation scheme described in Section~\ref{sec:repetition code} is efficient under two key conditions. First, decoding must be sufficiently fast to determine the Pauli frame of the data qubits prior to the application of the transversal $X^{1/4}$ gate. Second, the physical error rate must be low enough. Indeed, since the scheme relies on preselection, a high physical error rate can result in an excessively high average number of error correction rounds, rendering the protocol impractical. In this appendix, we introduce a modified version of the scheme from Section~\ref{sec:repetition code} that removes both of these constraints. The modified protocol, like the scheme presented in Section~\ref{sec:repetition code}, begins by initializing all data qubits in the $\ket{0}$ state and measuring the $X$-type stabilizers, but omits the preselection step. The transversal $X^{1/4}$ gate is then applied, and, unlike the original scheme, the $X$ stabilizers are measured again over multiple rounds.

\begin{figure}[t]
    \centering
    \includegraphics[width=0.48\textwidth]{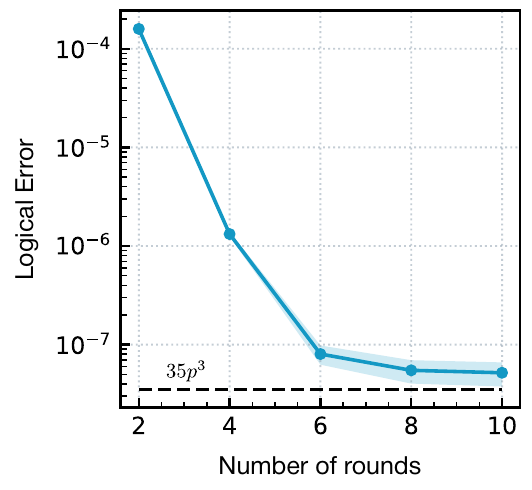}
    \caption{Logical error probability of a $|X^{1/4}\rangle$ magic state prepared using the unfolding distillation protocol presented in Appendix~\ref{app:Clifford ladder}. A circuit-level physical phase-flip error rate of $p_Z = 10^{-3}$ per operation is assumed. A minimum number of $X$-type stabilizer rounds is required before and after the application of the transversal $X^{1/4}$ gate to reach the expected distillation logical error rate of $35p^3$, where $p$ is the error probability of the $X^{-1/4}$ gate. This approach avoids preselection, and the total number of rounds must be multiplied by a factor of 3/2 compared to the simulation, as explained in Appendix~\ref{app:Clifford ladder}. Thus, a magic state with a logical error rate of $6 \times 10^{-8}$ can be prepared using 12 rounds of error correction. Appendix~\ref{app:high error} provides simulation results for larger phase-flip error rates.}
    \label{fig:clifford ladder}
\end{figure}

This modification offers three main benefits, with the potential trade-off of requiring more error correction rounds. First, although preselection is not performed before the $X^{-1/4}$ gate, errors on ancilla qubits in the stabilizer measurement round preceding the gate can be identified through subsequent rounds, and therefore most such errors do not contribute to the final distillation error. Second, if the decoding is too slow and an incorrect Pauli frame is used during the application of the $X^{\pm 1/4}$ gate, a corrective $X^{\pm 1/2}$ gate can be applied afterward to compensate. Third, because there is no preselection, the $X^{-1/4}$ gate can be applied immediately after the CNOT gate without waiting for the stabilizer measurement. This avoids additional idle time and explains why the magic state logical error is lower-bounded by $35(p^{\text{CNOT, target}}_Z + p^{X^{-1/4}}_X + p^{X^{-1/4}}_Y)^3$ rather than by $35(p^{\text{CNOT, target}}_Z + p^{\text{idle, mmt}}_{Z} + p^{X^{-1/4}}_X + p^{X^{-1/4}}_Y)^3$, where $p^{\text{CNOT, target}}_Z$ denotes the probability of a $Z$ error on the target (data) qubits of the CNOT gates prior to the $X^{-1/4}$ gates, $p^{\text{idle, mmt}}_{Z}$ represents the probability of a $Z$ error during the $X$ measurement of the final round of X-stabilizer measurements and $p^{X^{-1/4}}_X$ and $p^{X^{-1/4}}_Y$ are the probabilities of $X$ and $Y$ errors, respectively, on the $X^{-1/4}$ gate. If the physical error rate is sufficiently low, preselection can be performed before applying the $X^{\pm 1/2}$ corrections, followed directly by a measurement of the data qubits in the $Z$ basis. Equivalently, this could be performed with an $S/S^\dagger$ gate followed by a measurement in the $X$ basis. Since the $X^{\pm 1/2}$ and $S/S^\dagger$ are Clifford gates, the Pauli frame does not need to be determined in advance but can be tracked classically, removing the need for fast decoding. Alternatively, if the physical error rate is too high for preselection to succeed within a reasonable number of rounds, the preselection step can be omitted. In this case, the $X^{\pm 1/2}$ gate must be followed by additional rounds of $X$ stabilizer measurements. Once a sufficient number of rounds have been completed to ensure reliable detection of errors, all qubits can be measured in the $Z$ basis to conclude the protocol. We note that the corrective $X^{\pm 1/2}$ gates could introduce additional errors on top of those from the $X^{-1/4}$ gates. However, since such corrections occur infrequently, we choose to neglect their contribution to our error estimates.

In the simulations, we focus on the protocol without any preselection. Since we are restricted to Clifford simulations (see Appendix~\ref{app:simulation} for more details), we consider the following protocol. All data qubits are initialized in the $\ket{0}$ state, followed by several rounds of $X$-stabilizer measurements, the application of a transversal $X^{1/2}$ gate, followed by an equal number of subsequent rounds of $X$-stabilizer measurements. Figure~\ref{fig:clifford ladder} presents simulation results for $p_Z=10^{-3}$. When a sufficient number of error correction rounds is applied both before and after the transversal $X^{1/2}$ gate, the logical error rate reaches the expected $35p^3$ logical error for distillation, where $p$ is the physical error rate of the $X^{1/2}$ gate, without requiring any preselection. In a physical implementation, the protocol would require $3/2$ times as many rounds, as it would necessitate an initial transversal $X^{1/4}$ gate followed by corrective $X^{\pm 1/2}$ gates, each separated by a sufficient number of stabilizer rounds. Thus, the numbers reported in Table~\ref{tab:overhead} account for a round count increased by a factor of $3/2$ compared to the simulation results shown in Figure~\ref{fig:clifford ladder}. Simulations at higher error rates are presented in Appendix~\ref{app:high error}.

\section{Unfolded distillation with higher physical error rate}
\label{app:high error}

\begin{figure}[ht]
    \centering
    \includegraphics[width=0.48\textwidth]{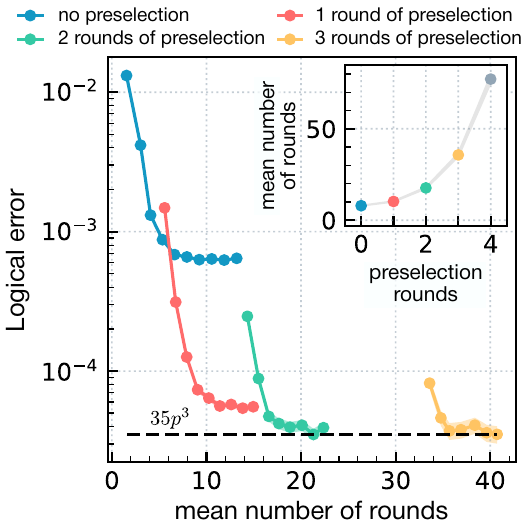}
    \caption{Logical error probability of the $|X^{1/4}\rangle$ magic state prepared using the unfolded distillation protocol. A circuit-level physical phase-flip error rate of $p_Z = 5 \times 10^{-3}$ per operation is assumed. Due to the noise in $X$-type stabilizer measurements, a minimum number of rounds is required to achieve the expected distillation logical error rate of $35p^3$, where $p$ is the error probability of the $X^{-1/4}$ gate. Preselection refers to repeatedly measuring the $X$-type stabilizers until, for a specified number of consecutive rounds, all outcomes are $+1$. The mean number of rounds includes both preselection and the final acceptance check based on the $Z$-type stabilizers of the Hadamard Reed-Muller code. At the assumed physical error rate, preselection becomes costly. As a result, two rounds of preselection and approximately 18 total rounds are required to reach the maximum achievable magic state fidelity. The inset shows how the mean number of rounds scales with the number of preselection rounds. This scaling is exponential, and at this level of physical noise, the scheme presented in Appendix~\ref{app:Clifford ladder} becomes preferable.}
    \label{fig: high error}
\end{figure}

Unfolded distillation, unlike schemes based on Clifford measurements, remains robust even at relatively high physical error rates. This is illustrated in Figure~\ref{fig: high error}, which shows the logical error rate of a magic state prepared with a physical error rate of $p_Z = 5 \times 10^{-3}$. At this error rate, the repetition code distance must be increased from $d=9$ to $d=11$ to avoid limiting the logical fidelity of the magic state. Despite the high error rate, the logical error continues to follow the expected $35p^3$ scaling behavior characteristic of distillation. However, due to the need for preselection, the number of rounds required for preparation increases. In this case, approximately 18 rounds are necessary.

\begin{figure}[ht]
    \centering
    \includegraphics[width=0.48\textwidth]{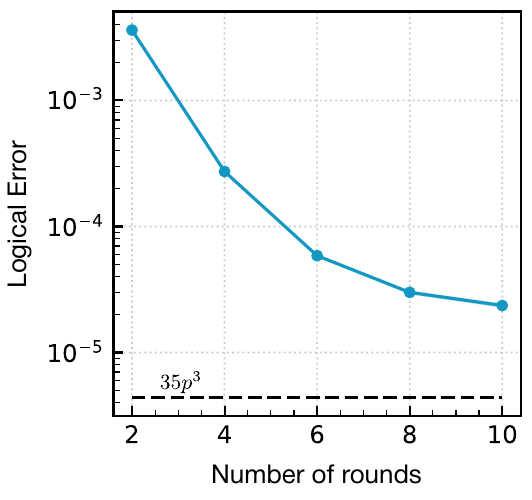}
    \caption{Logical error probability of a $|X^{1/4}\rangle$ magic state prepared using the unfolding distillation protocol presented in Appendix~\ref{app:Clifford ladder}. A circuit-level physical phase-flip error rate of $p_Z = 5 \times 10^{-3}$ per operation is assumed. A minimum number of $X$-type stabilizer rounds is required before and after the application of the transversal $X^{1/4}$ gate to reach a plateau in the logical error rate. This approach avoids preselection, and the total number of rounds must be multiplied by a factor of 3/2 compared to the simulation, as explained in Appendix~\ref{app:Clifford ladder}. Although higher-order errors prevent reaching the $35p^3$ bound set by distillation (see Appendix~\ref{app:high error}), where $p$ denotes the error rate of the $X^{-1/4}$ gate, a magic state with a logical error rate of $2 \times 10^{-5}$ can still be prepared using 15 rounds of error correction.}
    \label{fig:clifford ladder high error}
\end{figure}

In such high-error regimes, the alternative scheme described in Appendix~\ref{app:Clifford ladder} is more efficient, as it avoids preselection altogether. Figure~\ref{fig:clifford ladder high error} shows the resulting magic state fidelity under the same physical error rate $p_Z = 5 \times 10^{-3}$. Taking into account the additional factor of $3/2$ (see Appendix~\ref{app:Clifford ladder}), the total number of rounds is given by 15, and the scheme also achieves a slightly lower logical error rate compared to the scheme presented in Figure~\ref{fig: high error}. However, the logical error rate as a function of the number of rounds does not converge to the $35p^3$ limit set by distillation. Empirical analysis of failing configurations reveals that logical failures result from higher-weight errors (greater than three), involving both $X^{-1/4}$ gates and errors occurring during the error correction cycles before and after the transversal $X^{1/4}$ gate. Although such high-weight errors are negligible when $p_Z = 10^{-3}$, as shown in Figure~\ref{fig:clifford ladder}, they nonetheless become important at $p_Z = 5 \times 10^{-3}$.

\section{Simulating non-Clifford circuit}
\label{app:simulation}

All simulations were performed using the Clifford simulator Stim~\cite{gidney2021stim}. Since Stim cannot handle non-Clifford gates, the $X^{1/4}$ gates were replaced with $X^{1/2}$ gates, effectively simulating the distillation of $\ket{X^{1/2}}$ states instead of $\ket{X^{1/4}}$ states. This simplification is necessary for efficient simulation of the protocol and to probe low logical error rates. Importantly, the logical error rate obtained for $\ket{X^{1/2}}$ is an upper bound for that of $\ket{X^{1/4}}$. This is because, in addition to assuming equal gate error rates for $X^{1/4}$ and $X^{1/2}$, the two rotations differ in how they propagate Pauli errors. A $Z$ error occurring before an $X^{1/2}$ gate is conjugated to an $XZ$ error, whereas before an $X^{1/4}$ gate, it is conjugated to an $X^{1/2}Z$. Upon measuring the data qubit in the $Z$ basis, an $XZ$ error flips the measurement outcome deterministically, while an $X^{1/2}Z$ error flips it only with 50\% probability. Consequently, the contribution to the logical error rate from $Z$ errors immediately preceding the $X$ rotation is expected to be twice as large in the $X^{1/2}$ case as in the $X^{1/4}$ case.

Because a single $Z$ error on a data qubit can trigger up to four $X$-type stabilizers, standard matching algorithms are not applicable to decoding the unfolded code. Instead, all simulations presented in this work used the BP+OSD decoder~\cite{panteleev2021degenerate, roffe2020decoding, Roffe_LDPC_Python_tools_2022}. While this decoder is relatively slow, the modest size of the unfolded code could make it tractable in practice. It is conceivable that dedicated hardware, combined with recent improvements in decoding methods, could enable real-time decoding~\cite{muller2025improved, koutsioumpas2025automorphism,gong2024toward, ott2025decision, hillmann2024localized, wolanski2024ambiguity}. We also note that the scheme presented in Appendix~\ref{app:Clifford ladder} can tolerate a slower decoding speed.

\section{Search of the code}
\label{app:code search}

In this appendix, we detail how the code search is performed. The goal is to find a phase-flip code with a 2D nearest neighbor layout, but whose group of X stabilizers contains as a subgroup the group of X stabilizers of the Hadamard Reed-Muller code. As seen in Section~\ref{sec:repetition code}B, this is equivalent to finding a phase-flip code where each of the 15 $\pi/4$ $X$ rotations of the unencoded distillation circuit of Figure~\ref{fig:mapping XL}(b) can be performed by applying a $X^{1/4}$ gate on a single physical qubit. Thus, the goal is to find a code with 5 logical qubits where each logical operator of the set $\mathcal S _{RM} = \{ X_{L,i} \mid 1 \le i \le 5 \} \cup \{ X_{L,i} X_{L,j} X_{L,k} \mid 1 \le i, j, k \le 5, i \ne j \ne k \} \cup \{ X_{L,1} X_{L,2} X_{L,3} X_{L,4} X_{L,5} \}$ corresponds to a physical $X$ gate.

As the space of possible codes is too large for an exhaustive search, we restricted the search to codes where the parity check matrix $H$ can be put in the following form simply by numbering the qubits properly
\begin{equation}
H = \left( \begin{array}{c|c}
H'\ & \begin{matrix}
1 & 0 & 0 & \cdots & 0 \\
* & 1 & 0 & \cdots & 0 \\
* & * & 1 & \cdots & 0 \\
\vdots & \vdots & \vdots & \ddots & \vdots \\
* & * & * & \cdots & 1 \\
\end{matrix}
\end{array} \right)
\end{equation}
where $H' \in \mathbb{F}_2^{(n-k) \times k}$

If any parity check matrix can be put into this form by combining the rows of $H$, we here require that the parity check matrix representing the physical stabilizers that are measured is natively in this form. This ensures that all checks are independent, and thus that the number of logical qubits is the one required, here $k = 5$. This matrix form induces a natural numbering of the qubits that corresponds to the index of the column in $H$. The $k$ first qubits can be viewed as \textit{information} qubits as fixing their value fixes the codeword of the code. The $n-k$ remaining qubits can be viewed as \textit{redundancy} qubits. 

We choose the logical basis so that $X_{L,i} = X_i$ for the physical qubits $i \in \{1,...,k\}$. Then the logical operation corresponding to $X_j$ can be easily computed recursively for any redundancy qubit. For instance, if the qubit $j$ is involved in the parity check $X_a X_b X_c X_j$ with $a,b,c<j$, then $X_{j} = X_{L,a} X_{L,b} X_{L,c}$.

The restriction to such codes enables the search. The search is performed with the use of the Z3~\cite{z3} and Kissat~\cite{kissat,kissat-github} SAT solver. We start with a $L \times L$ square grid of data qubits. \textit{Connection} variables represent which qubit participates in each check. To each qubit is associated an \textit{ordering} variable representing its column index of the parity check matrix. A \textit{rotation} variable is associated with each qubit representing the logical rotation corresponding to a physical $X$ gate on this qubit.

We then impose the following constraints on the variables. Constraints are imposed on the \textit{connection} variables such that the parity checks are only between nearest neighbor qubits on the $L$ by $L$ grid. It follows that the weight of each parity check is at most 4. \textit{Rotation} variables are constrained such that $X_j = X_{L,j}$ for $j \le k$ and $X_j = X_{L,a} X_{L,b} X_{L,c}$ for $j > k$ if the qubit $j$ is involved in the parity check $X_j X_{a} X_{b} X_{c}$ with $a,b,c < j$. Finally, we impose that there exists a qubit with a \textit{rotation} variable corresponding to each element of the set $\mathcal S _{RM}$. The code corresponding to the search can be found at~\cite{CodeGithub}.

\bibliographystyle{unsrtnat}
\bibliography{biblio}

\begin{thebibliography}{81}
\providecommand{\natexlab}[1]{#1}
\providecommand{\url}[1]{\texttt{#1}}
\expandafter\ifx\csname urlstyle\endcsname\relax
  \providecommand{\doi}[1]{doi: #1}\else
  \providecommand{\doi}{doi: \begingroup \urlstyle{rm}\Url}\fi

\bibitem[Shor(1994)]{shor1994algorithms}
Peter~W Shor.
\newblock Algorithms for quantum computation: discrete logarithms and factoring.
\newblock In \emph{Proceedings 35th annual symposium on foundations of computer science}, pages 124--134. Ieee, 1994.

\bibitem[Dalzell et~al.(2023)Dalzell, McArdle, Berta, Bienias, Chen, Gily{\'e}n, Hann, Kastoryano, Khabiboulline, Kubica, et~al.]{dalzell2023quantum}
Alexander~M Dalzell, Sam McArdle, Mario Berta, Przemyslaw Bienias, Chi-Fang Chen, Andr{\'a}s Gily{\'e}n, Connor~T Hann, Michael~J Kastoryano, Emil~T Khabiboulline, Aleksander Kubica, et~al.
\newblock Quantum algorithms: A survey of applications and end-to-end complexities.
\newblock \emph{arXiv preprint arXiv:2310.03011}, 2023.

\bibitem[Beverland et~al.(2022)Beverland, Murali, Troyer, Svore, Hoefler, Kliuchnikov, Low, Soeken, Sundaram, and Vaschillo]{beverland2022assessing}
Michael~E Beverland, Prakash Murali, Matthias Troyer, Krysta~M Svore, Torsten Hoefler, Vadym Kliuchnikov, Guang~Hao Low, Mathias Soeken, Aarthi Sundaram, and Alexander Vaschillo.
\newblock Assessing requirements to scale to practical quantum advantage.
\newblock \emph{arXiv preprint arXiv:2211.07629}, 2022.

\bibitem[Acharya et~al.(2024)Acharya, Abanin, Aghababaie-Beni, Aleiner, Andersen, Ansmann, Arute, Arya, Asfaw, Astrakhantsev, et~al.]{acharya2024quantum}
Rajeev Acharya, Dmitry~A Abanin, Laleh Aghababaie-Beni, Igor Aleiner, Trond~I Andersen, Markus Ansmann, Frank Arute, Kunal Arya, Abraham Asfaw, Nikita Astrakhantsev, et~al.
\newblock Quantum error correction below the surface code threshold.
\newblock \emph{Nature}, 2024.

\bibitem[Moses et~al.(2023)Moses, Baldwin, Allman, Ancona, Ascarrunz, Barnes, Bartolotta, Bjork, Blanchard, Bohn, et~al.]{moses2023race}
Steven~A Moses, Charles~H Baldwin, Michael~S Allman, R~Ancona, L~Ascarrunz, C~Barnes, J~Bartolotta, B~Bjork, P~Blanchard, M~Bohn, et~al.
\newblock A race-track trapped-ion quantum processor.
\newblock \emph{Physical Review X}, 13\penalty0 (4):\penalty0 041052, 2023.

\bibitem[Bluvstein et~al.(2024)Bluvstein, Evered, Geim, Li, Zhou, Manovitz, Ebadi, Cain, Kalinowski, Hangleiter, et~al.]{bluvstein2024logical}
Dolev Bluvstein, Simon~J Evered, Alexandra~A Geim, Sophie~H Li, Hengyun Zhou, Tom Manovitz, Sepehr Ebadi, Madelyn Cain, Marcin Kalinowski, Dominik Hangleiter, et~al.
\newblock Logical quantum processor based on reconfigurable atom arrays.
\newblock \emph{Nature}, 626\penalty0 (7997):\penalty0 58--65, 2024.

\bibitem[Shor(1996)]{shor1996fault}
Peter~W Shor.
\newblock Fault-tolerant quantum computation.
\newblock In \emph{Proceedings of 37th conference on foundations of computer science}, pages 56--65. IEEE, 1996.

\bibitem[Shor(1995)]{shor1995scheme}
Peter~W Shor.
\newblock Scheme for reducing decoherence in quantum computer memory.
\newblock \emph{Physical review A}, 52\penalty0 (4):\penalty0 R2493, 1995.

\bibitem[Steane(1996{\natexlab{a}})]{steane1996multiple}
Andrew Steane.
\newblock Multiple-particle interference and quantum error correction.
\newblock \emph{Proceedings of the Royal Society of London. Series A: Mathematical, Physical and Engineering Sciences}, 452\penalty0 (1954):\penalty0 2551--2577, 1996{\natexlab{a}}.

\bibitem[Aharonov and Ben-Or(1997)]{aharonov1997fault}
Dorit Aharonov and Michael Ben-Or.
\newblock Fault-tolerant quantum computation with constant error.
\newblock In \emph{Proceedings of the twenty-ninth annual ACM symposium on Theory of computing}, pages 176--188, 1997.

\bibitem[Kitaev(2003)]{kitaev2003fault}
A~Yu Kitaev.
\newblock Fault-tolerant quantum computation by anyons.
\newblock \emph{Annals of physics}, 303\penalty0 (1):\penalty0 2--30, 2003.

\bibitem[Knill et~al.(1998)Knill, Laflamme, and Zurek]{knill1998resilient}
Emanuel Knill, Raymond Laflamme, and Wojciech~H Zurek.
\newblock Resilient quantum computation.
\newblock \emph{Science}, 279\penalty0 (5349):\penalty0 342--345, 1998.

\bibitem[Eastin and Knill(2009)]{eastin2009restrictions}
Bryan Eastin and Emanuel Knill.
\newblock Restrictions on transversal encoded quantum gate sets.
\newblock \emph{Physical review letters}, 102\penalty0 (11):\penalty0 110502, 2009.

\bibitem[Bravyi and K{\"o}nig(2013)]{bravyi2013classification}
Sergey Bravyi and Robert K{\"o}nig.
\newblock Classification of topologically protected gates for local stabilizer codes.
\newblock \emph{Physical review letters}, 110\penalty0 (17):\penalty0 170503, 2013.

\bibitem[Bravyi and Kitaev(2005)]{bravyi2005universal}
Sergey Bravyi and Alexei Kitaev.
\newblock Universal quantum computation with ideal clifford gates and noisy ancillas.
\newblock \emph{Physical Review A—Atomic, Molecular, and Optical Physics}, 71\penalty0 (2):\penalty0 022316, 2005.

\bibitem[Zhou et~al.(2000)Zhou, Leung, and Chuang]{zhou2000methodology}
Xinlan Zhou, Debbie~W Leung, and Isaac~L Chuang.
\newblock Methodology for quantum logic gate construction.
\newblock \emph{Physical Review A}, 62\penalty0 (5):\penalty0 052316, 2000.

\bibitem[Chamberland et~al.(2022)Chamberland, Noh, Arrangoiz-Arriola, Campbell, Hann, Iverson, Putterman, Bohdanowicz, Flammia, Keller, et~al.]{chamberland2022building}
Christopher Chamberland, Kyungjoo Noh, Patricio Arrangoiz-Arriola, Earl~T Campbell, Connor~T Hann, Joseph Iverson, Harald Putterman, Thomas~C Bohdanowicz, Steven~T Flammia, Andrew Keller, et~al.
\newblock Building a fault-tolerant quantum computer using concatenated cat codes.
\newblock \emph{PRX Quantum}, 3\penalty0 (1):\penalty0 010329, 2022.

\bibitem[Gouzien et~al.(2023)Gouzien, Ruiz, Le~R{\'e}gent, Guillaud, and Sangouard]{gouzien2023performance}
{\'E}lie Gouzien, Diego Ruiz, Francois-Marie Le~R{\'e}gent, J{\'e}r{\'e}mie Guillaud, and Nicolas Sangouard.
\newblock Performance analysis of a repetition cat code architecture: Computing 256-bit elliptic curve logarithm in 9 hours with 126 133 cat qubits.
\newblock \emph{Physical review letters}, 131\penalty0 (4):\penalty0 040602, 2023.

\bibitem[Litinski(2019{\natexlab{a}})]{litinski2019magic}
Daniel Litinski.
\newblock Magic state distillation: Not as costly as you think.
\newblock \emph{Quantum}, 3:\penalty0 205, 2019{\natexlab{a}}.

\bibitem[Gidney et~al.(2024{\natexlab{a}})Gidney, Shutty, and Jones]{gidney2024magic}
Craig Gidney, Noah Shutty, and Cody Jones.
\newblock Magic state cultivation: growing t states as cheap as cnot gates.
\newblock \emph{arXiv preprint arXiv:2409.17595}, 2024{\natexlab{a}}.

\bibitem[Guillaud and Mirrahimi(2021)]{guillaud2021error}
J{\'e}r{\'e}mie Guillaud and Mazyar Mirrahimi.
\newblock Error rates and resource overheads of repetition cat qubits.
\newblock \emph{Physical Review A}, 103\penalty0 (4):\penalty0 042413, 2021.

\bibitem[Gidney and Fowler(2019)]{gidney2019efficient}
Craig Gidney and Austin~G Fowler.
\newblock Efficient magic state factories with a catalyzed $|\text{CC}{Z}\rangle$ to $2|{T}\rangle$ transformation.
\newblock \emph{Quantum}, 3:\penalty0 135, 2019.

\bibitem[Gottesman and Chuang(1999)]{gottesman1999demonstrating}
Daniel Gottesman and Isaac~L Chuang.
\newblock Demonstrating the viability of universal quantum computation using teleportation and single-qubit operations.
\newblock \emph{Nature}, 402\penalty0 (6760):\penalty0 390--393, 1999.

\bibitem[Steinacker et~al.(2024)Steinacker, Stuyck, Lim, Tanttu, Feng, Nickl, Serrano, Candido, Cifuentes, Hudson, et~al.]{steinacker2024300}
Paul Steinacker, Nard~Dumoulin Stuyck, Wee~Han Lim, Tuomo Tanttu, MengKe Feng, Andreas Nickl, Santiago Serrano, Marco Candido, Jesus~D Cifuentes, Fay~E Hudson, et~al.
\newblock A 300 mm foundry silicon spin qubit unit cell exceeding 99\% fidelity in all operations.
\newblock \emph{arXiv preprint arXiv:2410.15590}, 2024.

\bibitem[Aliferis and Preskill(2008)]{aliferis2008fault}
Panos Aliferis and John Preskill.
\newblock Fault-tolerant quantum computation against biased noise.
\newblock \emph{Physical Review A—Atomic, Molecular, and Optical Physics}, 78\penalty0 (5):\penalty0 052331, 2008.

\bibitem[Taylor et~al.(2005)Taylor, Engel, D{\"u}r, Yacoby, Marcus, Zoller, and Lukin]{taylor2005fault}
JM~Taylor, H-A Engel, W~D{\"u}r, Amnon Yacoby, CM~Marcus, P~Zoller, and MD~Lukin.
\newblock Fault-tolerant architecture for quantum computation using electrically controlled semiconductor spins.
\newblock \emph{Nature Physics}, 1\penalty0 (3):\penalty0 177--183, 2005.

\bibitem[Brito et~al.(2008)Brito, DiVincenzo, Koch, and Steffen]{brito2008efficient}
Frederico Brito, David~P DiVincenzo, Roger~H Koch, and Matthias Steffen.
\newblock Efficient one-and two-qubit pulsed gates for an oscillator-stabilized josephson qubit.
\newblock \emph{New Journal of Physics}, 10\penalty0 (3):\penalty0 033027, 2008.

\bibitem[Webster et~al.(2015)Webster, Bartlett, and Poulin]{webster2015reducing}
Paul Webster, Stephen~D Bartlett, and David Poulin.
\newblock Reducing the overhead for quantum computation when noise is biased.
\newblock \emph{Physical Review A}, 92\penalty0 (6):\penalty0 062309, 2015.

\bibitem[Mirrahimi et~al.(2014)Mirrahimi, Leghtas, Albert, Touzard, Schoelkopf, Jiang, and Devoret]{mirrahimi2014dynamically}
Mazyar Mirrahimi, Zaki Leghtas, Victor~V Albert, Steven Touzard, Robert~J Schoelkopf, Liang Jiang, and Michel~H Devoret.
\newblock Dynamically protected cat-qubits: a new paradigm for universal quantum computation.
\newblock \emph{New Journal of Physics}, 16\penalty0 (4):\penalty0 045014, 2014.

\bibitem[Guillaud and Mirrahimi(2019)]{guillaud2019repetition}
J{\'e}r{\'e}mie Guillaud and Mazyar Mirrahimi.
\newblock Repetition cat qubits for fault-tolerant quantum computation.
\newblock \emph{Physical Review X}, 9\penalty0 (4):\penalty0 041053, 2019.

\bibitem[Puri et~al.(2020)Puri, St-Jean, Gross, Grimm, Frattini, Iyer, Krishna, Touzard, Jiang, Blais, et~al.]{puri2020bias}
Shruti Puri, Lucas St-Jean, Jonathan~A Gross, Alexander Grimm, Nicholas~E Frattini, Pavithran~S Iyer, Anirudh Krishna, Steven Touzard, Liang Jiang, Alexandre Blais, et~al.
\newblock Bias-preserving gates with stabilized cat qubits.
\newblock \emph{Science advances}, 6\penalty0 (34):\penalty0 eaay5901, 2020.

\bibitem[R{\'e}glade et~al.(2024)R{\'e}glade, Bocquet, Gautier, Cohen, Marquet, Albertinale, Pankratova, Hall{\'e}n, Rautschke, Sellem, et~al.]{reglade2024quantum}
Ulysse R{\'e}glade, Adrien Bocquet, Ronan Gautier, Joachim Cohen, Antoine Marquet, Emanuele Albertinale, Natalia Pankratova, Mattis Hall{\'e}n, Felix Rautschke, L-A Sellem, et~al.
\newblock Quantum control of a cat qubit with bit-flip times exceeding ten seconds.
\newblock \emph{Nature}, 629\penalty0 (8013):\penalty0 778--783, 2024.

\bibitem[Rousseau et~al.(2025)Rousseau, Ruiz, Albertinale, d'Avezac, Banys, Blandin, Bourdaud, Campanaro, Cardoso, Cottet, et~al.]{rousseau2025enhancing}
R{\'e}mi Rousseau, Diego Ruiz, Emanuele Albertinale, Pol d'Avezac, Danielius Banys, Ugo Blandin, Nicolas Bourdaud, Giulio Campanaro, Gil Cardoso, Nathanael Cottet, et~al.
\newblock Enhancing dissipative cat qubit protection by squeezing.
\newblock \emph{arXiv preprint arXiv:2502.07892}, 2025.

\bibitem[Rojkov et~al.(2024)Rojkov, Simoni, Zapusek, Reiter, and Home]{rojkov2024stabilization}
Ivan Rojkov, Matteo Simoni, Elias Zapusek, Florentin Reiter, and Jonathan Home.
\newblock Stabilization of cat-state manifolds using nonlinear reservoir engineering.
\newblock \emph{arXiv preprint arXiv:2407.18087}, 2024.

\bibitem[Omanakuttan et~al.(2024)Omanakuttan, Buchemmavari, Gross, Deutsch, and Marvian]{omanakuttan2024fault}
Sivaprasad Omanakuttan, Vikas Buchemmavari, Jonathan~A Gross, Ivan~H Deutsch, and Milad Marvian.
\newblock Fault-tolerant quantum computation using large spin-cat codes.
\newblock \emph{PRX Quantum}, 5\penalty0 (2):\penalty0 020355, 2024.

\bibitem[DeBry et~al.(2025)DeBry, Meister, Martinez, Bruzewicz, Shi, Reens, McConnell, Chuang, and Chiaverini]{debry2025error}
Kyle DeBry, Nadine Meister, Agustin~Valdes Martinez, Colin~D Bruzewicz, Xiaoyang Shi, David Reens, Robert McConnell, Isaac~L Chuang, and John Chiaverini.
\newblock Error correction of a logical qubit encoded in a single atomic ion.
\newblock \emph{arXiv preprint arXiv:2503.13908}, 2025.

\bibitem[Ruiz et~al.(2025)Ruiz, Guillaud, Leverrier, Mirrahimi, and Vuillot]{ruiz2025ldpc}
Diego Ruiz, J{\'e}r{\'e}mie Guillaud, Anthony Leverrier, Mazyar Mirrahimi, and Christophe Vuillot.
\newblock Ldpc-cat codes for low-overhead quantum computing in 2d.
\newblock \emph{Nature Communications}, 16\penalty0 (1):\penalty0 1040, 2025.

\bibitem[Putterman et~al.(2025)Putterman, Noh, Hann, MacCabe, Aghaeimeibodi, Patel, Lee, Jones, Moradinejad, Rodriguez, et~al.]{putterman2025hardware}
Harald Putterman, Kyungjoo Noh, Connor~T Hann, Gregory~S MacCabe, Shahriar Aghaeimeibodi, Rishi~N Patel, Menyoung Lee, William~M Jones, Hesam Moradinejad, Roberto Rodriguez, et~al.
\newblock Hardware-efficient quantum error correction via concatenated bosonic qubits.
\newblock \emph{Nature}, 638\penalty0 (8052):\penalty0 927--934, 2025.

\bibitem[Hann et~al.(2024)Hann, Noh, Putterman, Matheny, Iverson, Fang, Chamberland, Painter, and Brand{\~a}o]{hann2024hybrid}
Connor~T Hann, Kyungjoo Noh, Harald Putterman, Matthew~H Matheny, Joseph~K Iverson, Michael~T Fang, Christopher Chamberland, Oskar Painter, and Fernando~GSL Brand{\~a}o.
\newblock Hybrid cat-transmon architecture for scalable, hardware-efficient quantum error correction.
\newblock \emph{arXiv preprint arXiv:2410.23363}, 2024.

\bibitem[Bonilla~Ataides et~al.(2021)Bonilla~Ataides, Tuckett, Bartlett, Flammia, and Brown]{bonilla2021xzzx}
J~Pablo Bonilla~Ataides, David~K Tuckett, Stephen~D Bartlett, Steven~T Flammia, and Benjamin~J Brown.
\newblock The xzzx surface code.
\newblock \emph{Nature communications}, 12\penalty0 (1):\penalty0 2172, 2021.

\bibitem[Singh et~al.(2022)Singh, Darmawan, Brown, and Puri]{singh2022high}
Shraddha Singh, Andrew~S Darmawan, Benjamin~J Brown, and Shruti Puri.
\newblock High-fidelity magic-state preparation with a biased-noise architecture.
\newblock \emph{Physical Review A}, 105\penalty0 (5):\penalty0 052410, 2022.

\bibitem[Steane(2002)]{steane2002quantum}
Andrew~M Steane.
\newblock Quantum reed-muller codes.
\newblock \emph{IEEE Transactions on Information Theory}, 45\penalty0 (5):\penalty0 1701--1703, 2002.

\bibitem[Koutsioumpas et~al.(2022)Koutsioumpas, Banfield, and Kay]{koutsioumpas2022smallest}
Stergios Koutsioumpas, Darren Banfield, and Alastair Kay.
\newblock The smallest code with transversal t.
\newblock \emph{arXiv preprint arXiv:2210.14066}, 2022.

\bibitem[Li(2015)]{li2015magic}
Ying Li.
\newblock A magic state’s fidelity can be superior to the operations that created it.
\newblock \emph{New Journal of Physics}, 17\penalty0 (2):\penalty0 023037, 2015.

\bibitem[Gidney(2023)]{gidney2023cleaner}
Craig Gidney.
\newblock Cleaner magic states with hook injection.
\newblock \emph{arXiv preprint arXiv:2302.12292}, 2023.

\bibitem[Fowler et~al.(2012)Fowler, Mariantoni, Martinis, and Cleland]{fowler2012surface}
Austin~G Fowler, Matteo Mariantoni, John~M Martinis, and Andrew~N Cleland.
\newblock Surface codes: Towards practical large-scale quantum computation.
\newblock \emph{Physical Review A—Atomic, Molecular, and Optical Physics}, 86\penalty0 (3):\penalty0 032324, 2012.

\bibitem[Haah and Hastings(2018)]{haah2018codes}
Jeongwan Haah and Matthew~B Hastings.
\newblock Codes and protocols for distilling $ t $, controlled-$ s $, and toffoli gates.
\newblock \emph{Quantum}, 2:\penalty0 71, 2018.

\bibitem[Horsman et~al.(2012)Horsman, Fowler, Devitt, and Van~Meter]{horsman2012surface}
Dominic Horsman, Austin~G Fowler, Simon Devitt, and Rodney Van~Meter.
\newblock Surface code quantum computing by lattice surgery.
\newblock \emph{New Journal of Physics}, 14\penalty0 (12):\penalty0 123011, 2012.

\bibitem[Fowler and Gidney(2018)]{fowler2018low}
Austin~G Fowler and Craig Gidney.
\newblock Low overhead quantum computation using lattice surgery.
\newblock \emph{arXiv preprint arXiv:1808.06709}, 2018.

\bibitem[Litinski(2019{\natexlab{b}})]{litinski2019game}
Daniel Litinski.
\newblock A game of surface codes: Large-scale quantum computing with lattice surgery.
\newblock \emph{Quantum}, 3:\penalty0 128, 2019{\natexlab{b}}.

\bibitem[Lee et~al.(2024)Lee, Thomsen, Fazio, Brown, and Bartlett]{lee2024low}
Seok-Hyung Lee, Felix Thomsen, Nicholas Fazio, Benjamin~J Brown, and Stephen~D Bartlett.
\newblock Low-overhead magic state distillation with color codes.
\newblock \emph{arXiv preprint arXiv:2409.07707}, 2024.

\bibitem[Lechner et~al.(2015)Lechner, Hauke, and Zoller]{lechner2015quantum}
Wolfgang Lechner, Philipp Hauke, and Peter Zoller.
\newblock A quantum annealing architecture with all-to-all connectivity from local interactions.
\newblock \emph{Science advances}, 1\penalty0 (9):\penalty0 e1500838, 2015.

\bibitem[Chao and Reichardt(2018)]{chao2018quantum}
Rui Chao and Ben~W Reichardt.
\newblock Quantum error correction with only two extra qubits.
\newblock \emph{Physical review letters}, 121\penalty0 (5):\penalty0 050502, 2018.

\bibitem[Gidney et~al.(2024{\natexlab{b}})Gidney, Newman, Brooks, and Jones]{gidney2024yoked}
Craig Gidney, Michael Newman, Peter Brooks, and Cody Jones.
\newblock Yoked surface codes.
\newblock \emph{arXiv preprint arXiv:2312.04522}, 2024{\natexlab{b}}.

\bibitem[Brown et~al.(2017)Brown, Laubscher, Kesselring, and Wootton]{brown2017poking}
Benjamin~J Brown, Katharina Laubscher, Markus~S Kesselring, and James~R Wootton.
\newblock Poking holes and cutting corners to achieve clifford gates with the surface code.
\newblock \emph{Physical Review X}, 7\penalty0 (2):\penalty0 021029, 2017.

\bibitem[Mantri et~al.(2017)Mantri, Demarie, and Fitzsimons]{mantri2017universality}
Atul Mantri, Tommaso~F Demarie, and Joseph~F Fitzsimons.
\newblock Universality of quantum computation with cluster states and (x, y)-plane measurements.
\newblock \emph{Scientific reports}, 7\penalty0 (1):\penalty0 42861, 2017.

\bibitem[Bombin and Martin-Delgado(2006)]{bombin2006topological}
Hector Bombin and Miguel~Angel Martin-Delgado.
\newblock Topological quantum distillation.
\newblock \emph{Physical review letters}, 97\penalty0 (18):\penalty0 180501, 2006.

\bibitem[Steane(1996{\natexlab{b}})]{steane1996error}
Andrew~M Steane.
\newblock Error correcting codes in quantum theory.
\newblock \emph{Physical Review Letters}, 77\penalty0 (5):\penalty0 793, 1996{\natexlab{b}}.

\bibitem[Kubica et~al.(2015)Kubica, Yoshida, and Pastawski]{kubica2015unfolding}
Aleksander Kubica, Beni Yoshida, and Fernando Pastawski.
\newblock Unfolding the color code.
\newblock \emph{New Journal of Physics}, 17\penalty0 (8):\penalty0 083026, 2015.

\bibitem[Selinger(2013)]{selinger2013quantum}
Peter Selinger.
\newblock Quantum circuits of t-depth one.
\newblock \emph{Physical Review A—Atomic, Molecular, and Optical Physics}, 87\penalty0 (4):\penalty0 042302, 2013.

\bibitem[Albert et~al.(2016)Albert, Shu, Krastanov, Shen, Liu, Yang, Schoelkopf, Mirrahimi, Devoret, and Jiang]{albert2016holonomic}
Victor~V Albert, Chi Shu, Stefan Krastanov, Chao Shen, Ren-Bao Liu, Zhen-Biao Yang, Robert~J Schoelkopf, Mazyar Mirrahimi, Michel~H Devoret, and Liang Jiang.
\newblock Holonomic quantum control with continuous variable systems.
\newblock \emph{Physical review letters}, 116\penalty0 (14):\penalty0 140502, 2016.

\bibitem[Heeres et~al.(2015)Heeres, Vlastakis, Holland, Krastanov, Albert, Frunzio, Jiang, and Schoelkopf]{heeres2015cavity}
Reinier~W Heeres, Brian Vlastakis, Eric Holland, Stefan Krastanov, Victor~V Albert, Luigi Frunzio, Liang Jiang, and Robert~J Schoelkopf.
\newblock Cavity state manipulation using photon-number selective phase gates.
\newblock \emph{Physical review letters}, 115\penalty0 (13):\penalty0 137002, 2015.

\bibitem[Krastanov et~al.(2015)Krastanov, Albert, Shen, Zou, Heeres, Vlastakis, Schoelkopf, and Jiang]{krastanov2015universal}
Stefan Krastanov, Victor~V Albert, Chao Shen, Chang-Ling Zou, Reinier~W Heeres, Brian Vlastakis, Robert~J Schoelkopf, and Liang Jiang.
\newblock Universal control of an oscillator with dispersive coupling to a qubit.
\newblock \emph{Physical Review A}, 92\penalty0 (4):\penalty0 040303, 2015.

\bibitem[Heeres et~al.(2017)Heeres, Reinhold, Ofek, Frunzio, Jiang, Devoret, and Schoelkopf]{heeres2017implementing}
Reinier~W Heeres, Philip Reinhold, Nissim Ofek, Luigi Frunzio, Liang Jiang, Michel~H Devoret, and Robert~J Schoelkopf.
\newblock Implementing a universal gate set on a logical qubit encoded in an oscillator.
\newblock \emph{Nature communications}, 8\penalty0 (1):\penalty0 94, 2017.

\bibitem[Eickbusch et~al.(2022)Eickbusch, Sivak, Ding, Elder, Jha, Venkatraman, Royer, Girvin, Schoelkopf, and Devoret]{eickbusch2022fast}
Alec Eickbusch, Volodymyr Sivak, Andy~Z Ding, Salvatore~S Elder, Shantanu~R Jha, Jayameenakshi Venkatraman, Baptiste Royer, Steven~M Girvin, Robert~J Schoelkopf, and Michel~H Devoret.
\newblock Fast universal control of an oscillator with weak dispersive coupling to a qubit.
\newblock \emph{Nature Physics}, 18\penalty0 (12):\penalty0 1464--1469, 2022.

\bibitem[Mehta et~al.(2025)Mehta, Teoh, Noh, Agrawal, Chamberlain, Chien, Curtis, Elfeky, Farzaneh, Gudlewski, et~al.]{mehta2025bias}
Nitish Mehta, James~D Teoh, Taewan Noh, Ankur Agrawal, Richard Chamberlain, Tzu-Chiao Chien, Jacob~C Curtis, Bassel~Heiba Elfeky, SM~Farzaneh, Benjamin Gudlewski, et~al.
\newblock Bias-preserving and error-detectable entangling operations in a superconducting dual-rail system.
\newblock \emph{arXiv preprint arXiv:2503.10935}, 2025.

\bibitem[Cod()]{CodeGithub}
Code is available at \url{https://github.com/DiegoRuiz-Git/Unfolded_distillation}, Diego Ruiz, 2025.

\bibitem[Gidney et~al.(2024{\natexlab{c}})Gidney, Jones, and Shutty]{gidney2024magicZenodo}
Craig Gidney, Cody Jones, and Nicholas Shutty.
\newblock {Data for "Magic state cultivation: growing T states as cheap as CNOT gates"}, 2024{\natexlab{c}}.
\newblock URL \url{https://doi.org/10.5281/zenodo.13777072}.

\bibitem[Gidney(2021)]{gidney2021stim}
Craig Gidney.
\newblock Stim: a fast stabilizer circuit simulator.
\newblock \emph{{Quantum}}, 5:\penalty0 497, July 2021.
\newblock ISSN 2521-327X.
\newblock \doi{10.22331/q-2021-07-06-497}.
\newblock URL \url{https://doi.org/10.22331/q-2021-07-06-497}.

\bibitem[Panteleev and Kalachev(2021)]{panteleev2021degenerate}
Pavel Panteleev and Gleb Kalachev.
\newblock Degenerate quantum ldpc codes with good finite length performance.
\newblock \emph{Quantum}, 5:\penalty0 585, 2021.

\bibitem[Roffe et~al.(2020)Roffe, White, Burton, and Campbell]{roffe2020decoding}
Joschka Roffe, David~R. White, Simon Burton, and Earl Campbell.
\newblock Decoding across the quantum low-density parity-check code landscape.
\newblock \emph{Physical Review Research}, 2\penalty0 (4), Dec 2020.
\newblock ISSN 2643-1564.
\newblock \doi{10.1103/physrevresearch.2.043423}.
\newblock URL \url{http://dx.doi.org/10.1103/PhysRevResearch.2.043423}.

\bibitem[Roffe(2022)]{Roffe_LDPC_Python_tools_2022}
Joschka Roffe.
\newblock {LDPC: Python tools for low density parity check codes}, 2022.
\newblock URL \url{https://pypi.org/project/ldpc/}.

\bibitem[M{\"u}ller et~al.(2025)M{\"u}ller, Alexander, Beverland, B{\"u}hler, Johnson, Maurer, and Vandeth]{muller2025improved}
Tristan M{\"u}ller, Thomas Alexander, Michael~E Beverland, Markus B{\"u}hler, Blake~R Johnson, Thilo Maurer, and Drew Vandeth.
\newblock Improved belief propagation is sufficient for real-time decoding of quantum memory.
\newblock \emph{arXiv preprint arXiv:2506.01779}, 2025.

\bibitem[Koutsioumpas et~al.(2025)Koutsioumpas, Sayginel, Webster, and Browne]{koutsioumpas2025automorphism}
Stergios Koutsioumpas, Hasan Sayginel, Mark Webster, and Dan~E Browne.
\newblock Automorphism ensemble decoding of quantum ldpc codes.
\newblock \emph{arXiv preprint arXiv:2503.01738}, 2025.

\bibitem[Gong et~al.(2024)Gong, Cammerer, and Renes]{gong2024toward}
Anqi Gong, Sebastian Cammerer, and Joseph~M Renes.
\newblock Toward low-latency iterative decoding of qldpc codes under circuit-level noise.
\newblock \emph{arXiv preprint arXiv:2403.18901}, 2024.

\bibitem[Ott et~al.(2025)Ott, Het{\'e}nyi, and Beverland]{ott2025decision}
KR~Ott, B~Het{\'e}nyi, and ME~Beverland.
\newblock Decision-tree decoders for general quantum ldpc codes, feb.
\newblock \emph{arXiv preprint arXiv:2502.16408}, 2025.

\bibitem[Hillmann et~al.(2024)Hillmann, Berent, Quintavalle, Eisert, Wille, and Roffe]{hillmann2024localized}
Timo Hillmann, Lucas Berent, Armanda~O Quintavalle, Jens Eisert, Robert Wille, and Joschka Roffe.
\newblock Localized statistics decoding: A parallel decoding algorithm for quantum low-density parity-check codes.
\newblock \emph{arXiv preprint arXiv:2406.18655}, 2024.

\bibitem[Wolanski and Barber(2024)]{wolanski2024ambiguity}
Stasiu Wolanski and Ben Barber.
\newblock Ambiguity clustering: an accurate and efficient decoder for qldpc codes.
\newblock \emph{arXiv preprint arXiv:2406.14527}, 2024.

\bibitem[Bjorner and et~al()]{z3}
Nikolaj Bjorner and et~al.
\newblock Z3 theorem prover.
\newblock URL \url{https://github.com/Z3Prover/z3}.

\bibitem[Biere et~al.(2024)Biere, Faller, Fazekas, Fleury, Froleyks, and Pollitt]{kissat}
Armin Biere, Tobias Faller, Katalin Fazekas, Mathias Fleury, Nils Froleyks, and Florian Pollitt.
\newblock {CaDiCaL}, {Gimsatul}, {IsaSAT} and {Kissat} entering the {SAT Competition 2024}.
\newblock In Marijn Heule, Markus Iser, Matti J{\"a}rvisalo, and Martin Suda, editors, \emph{Proc.~of {SAT Competition} 2024 -- Solver, Benchmark and Proof Checker Descriptions}, volume B-2024-1 of \emph{Department of Computer Science Report Series B}, pages 8--10. University of Helsinki, 2024.

\bibitem[Biere et~al.()Biere, Faller, Fazekas, Fleury, Froleyks, and Pollitt]{kissat-github}
Armin Biere, Tobias Faller, Katalin Fazekas, Mathias Fleury, Nils Froleyks, and Florian Pollitt.
\newblock The kissat sat solver.
\newblock URL \url{https://github.com/arminbiere/kissat}.

\end{thebibliography}

\end{document}